\documentclass{article} %
\usepackage{colm2024_conference} %

\usepackage{microtype}
\usepackage[hidelinks]{hyperref} %
\usepackage{url}
\usepackage{booktabs}
\usepackage{xspace}
\usepackage{enumitem}
\usepackage[pdftex]{graphicx}
\usepackage[flushleft]{threeparttable}
\usepackage{makecell}
\usepackage{multirow}
\usepackage{xcolor, colortbl}
\usepackage{subcaption}
\usepackage{tikz}
\usepackage{wrapfig}
\usepackage{textcomp}  %
\usepackage{scalerel}  %
\usepackage{tipa}
\usepackage{pifont}
\usepackage[most]{tcolorbox}
\usepackage{listings}
\usepackage{caption}
\newcommand{\dataset}{\textsc{EvoEval}\xspace} %
\newcommand{\tech}{\textsc{EvoEval}\xspace}

\newcommand{\humaneval}{\textsc{HumanEval}\xspace}
\newcommand{\evalplus}{\textsc{EvalPlus}\xspace}
\newcommand{\apps}{{APPS}\xspace}
\newcommand{\mbpp}{{MBPP}\xspace}

\newcommand{\eedifficult}{\textsc{Difficult}\xspace}
\newcommand{\eecreative}{\textsc{Creative}\xspace}
\newcommand{\eesubtle}{\textsc{Subtle}\xspace}
\newcommand{\eecombine}{\textsc{Combine}\xspace}
\newcommand{\eetoolusing}{\textsc{Tool\_Use}\xspace}
\newcommand{\eeverbose}{\textsc{Verbose}\xspace}
\newcommand{\eeconcise}{\textsc{Concise}\xspace}
\newcommand{\eedecompose}{\textsc{Decompose}\xspace}

\newcommand{\totalmodels}{51\xspace}
\newcommand{\difficultdecrease}{58.7\%\xspace}
\newcommand{\creativedecrease}{50.2\%\xspace}
\newcommand{\subtledecrease}{5.0\%\xspace}
\newcommand{\combinedecrease}{78.1\%\xspace}
\newcommand{\toolusingdecrease}{4.9\%\xspace}
\newcommand{\subtlecomparisondecrease}{24.0\%\xspace}

\newcommand{\totaldecrease}{39.4\%\xspace}
\newcommand{\maxdecrease}{47.7\%\xspace}
\newcommand{\mindecrease}{19.6\%\xspace}

\newcommand{\codegen}{{CodeGen}\xspace}
\newcommand{\starcoder}{{StarCoder}\xspace}

\newcommand{\codex}{\textsc{Codex}\xspace}

\newcommand{\gptfour}{GPT-4\xspace}
\newcommand{\gemini}{Gemini\xspace}
\newcommand{\claude}{Claude\xspace}
\newcommand{\chatgpt}{ChatGPT\xspace}

\newcommand{\codellama}{CodeLlama\xspace}
\newcommand{\mistral}{Mistral\xspace}

\newcommand{\wizardcoder}{{WizardCoder}\xspace}
\newcommand{\deepseekinstruct}{DeepSeeker-Inst\xspace}
\newcommand{\deepseek}{DeepSeeker\xspace}
\newcommand{\deepseekvonefive}{DeepSeeker-1.5\xspace}
\newcommand{\codellamainstruct}{CodeLlama-Inst\xspace}
\newcommand{\phitwo}{Phi-2\xspace}
\newcommand{\openchat}{OpenChat\xspace}
\newcommand{\qwen}{Qwen-1.5\xspace}
\newcommand{\qwenb}{Qwen\xspace}
\newcommand{\palm}{PaLM-2\xspace}
\newcommand{\phindllamatwo}{Phind-CodeLlama-2\xspace}
\newcommand{\mistralinstruct}{Mistral-Inst\xspace}
\newcommand{\mixtralinstruct}{Mixtral-Inst\xspace}
\newcommand{\codemillenials}{Code Millenials\xspace}
\newcommand{\xwincoder}{XwinCoder\xspace}
\newcommand{\stablecode}{stable-code\xspace}
\newcommand{\gemma}{Gemma\xspace}
\newcommand{\speechlesscodellama}{Speechless-CL\xspace}
\newcommand{\starcodertwo}{StarCoder2\xspace}
\newcommand{\magicoder}{Magicoder\xspace}

\newcommand{\llmfull}{large language model\xspace}
\newcommand{\llm}{LLM\xspace}

\newcommand{\passat}[1]{\textls[-25]{pass{@}\(#1\)}\xspace}
\newcommand{\parabf}[1]{\vspace{.03in}\noindent \textbf{#1}}
\newcommand{\CodeIn}[1]{{\small \texttt{#1}}}
\newcommand{\Comment}[1]{}

\newcommand{\eg}{e.g.,\xspace}
\newcommand{\ie}{i.e.,\xspace}

\newcommand*\circled[1]{\scalebox{0.8}{\tikz[baseline=(char.base)]{
\node[anchor=text, shape=circle,fill, inner sep=0pt, minimum size=1.2em] (char) {\footnotesize \textcolor{white}{#1}};}}}

\newcommand{\speech}{\scalerel*{\includegraphics{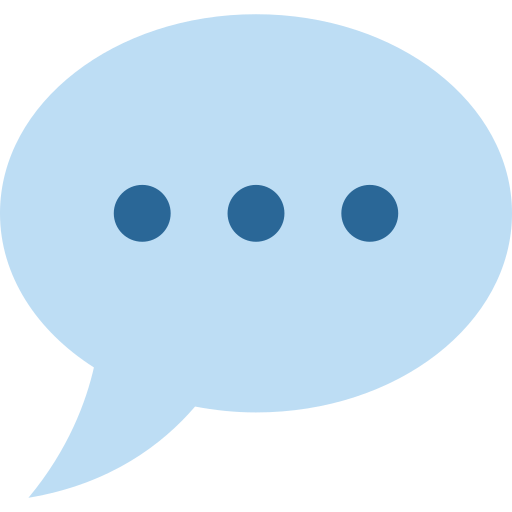}}{\textrm{C}}\xspace}

\newcommand{\openai}{\scalerel*{\includegraphics{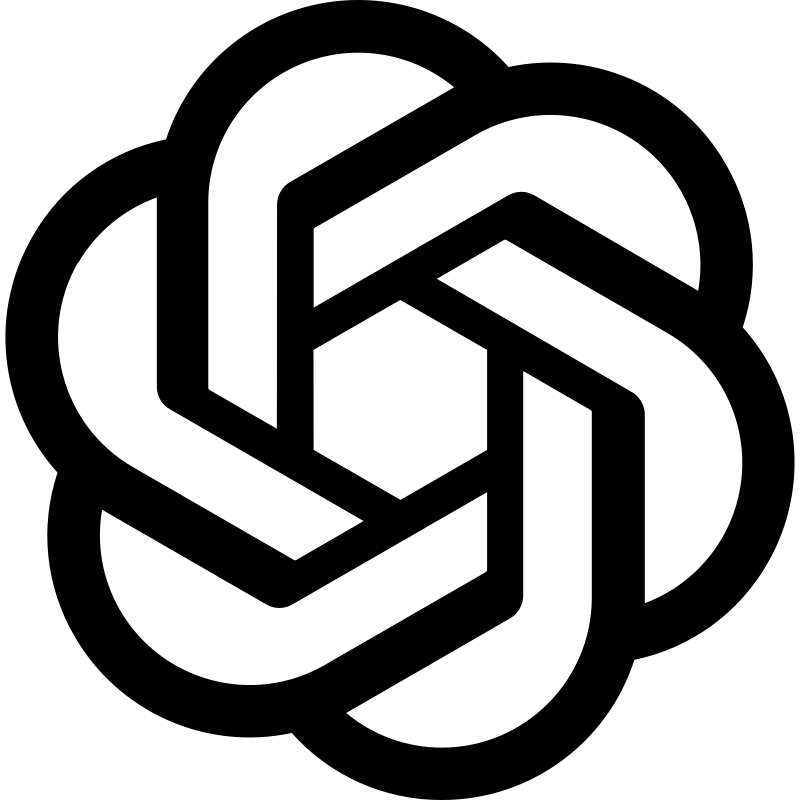}}{\textrm{C}}\xspace}
\newcommand{\google}{\scalerel*{\includegraphics{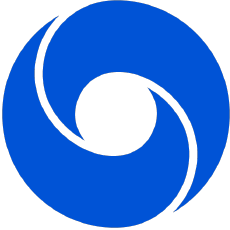}}{\textrm{C}}\xspace}
\newcommand{\meta}{\scalebox{0.7}{\scalerel*{\includegraphics{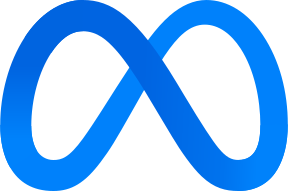}}{\textrm{C}}}\xspace}
\newcommand{\anthropic}{\scalebox{1}{\scalerel*{\includegraphics{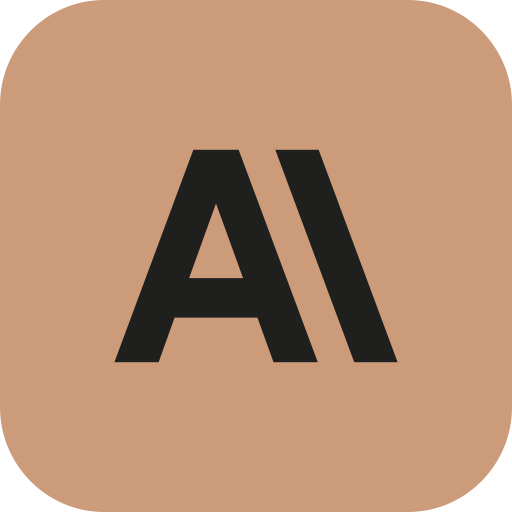}}{\textrm{C}}}\xspace}
\newcommand{\deepseeklogo}{\scalebox{0.7}{\scalerel*{\includegraphics{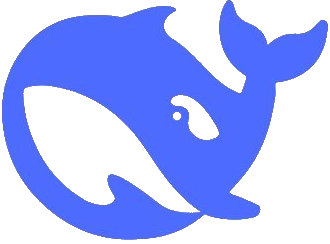}}{\textrm{C}}}\xspace}
\newcommand{\mistrallogo}{\scalebox{1}{\scalerel*{\includegraphics{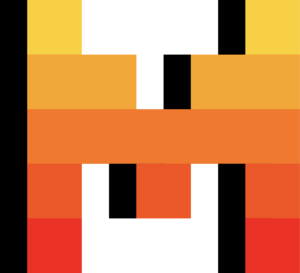}}{\textrm{C}}}\xspace}
\newcommand{\microsoft}{\scalebox{1}{\scalerel*{\includegraphics{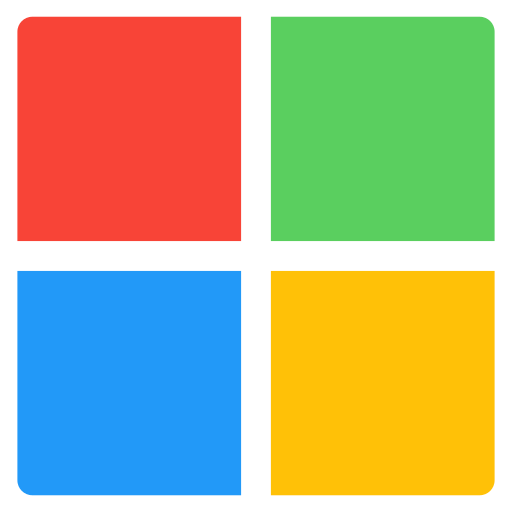}}{\textrm{C}}}\xspace}

\newcommand{\uiuc}{\scalebox{1}{\scalerel*{\includegraphics{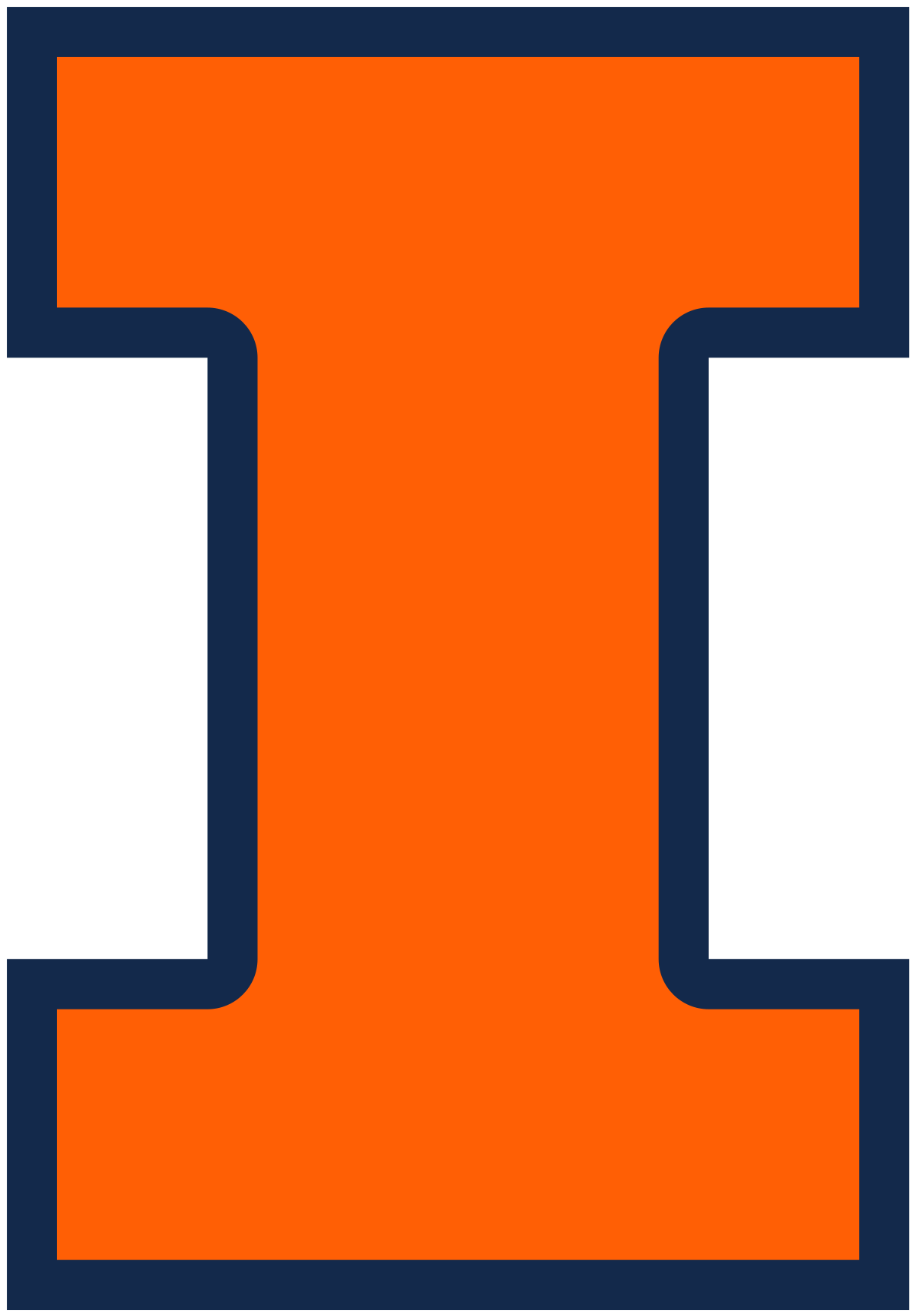}}{\textrm{C}}}\xspace}
\newcommand{\butterfly}{\scalerel*{\includegraphics{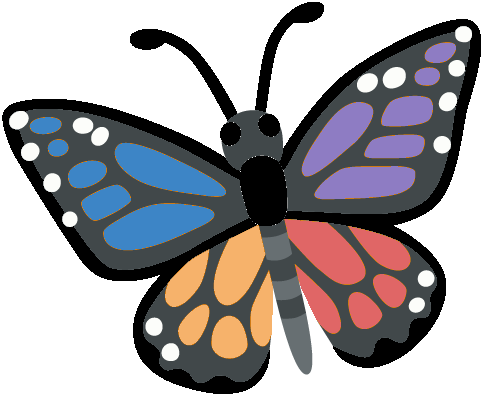}}{\textrm{\textbigcircle}}\xspace}
\newcommand{\butterflysmall}{\scalerel*{\includegraphics{resources/butterfly_dark.png}}{\textrm{C}}\xspace}

\newcommand{\butterflyblue}{\scalerel*{\includegraphics{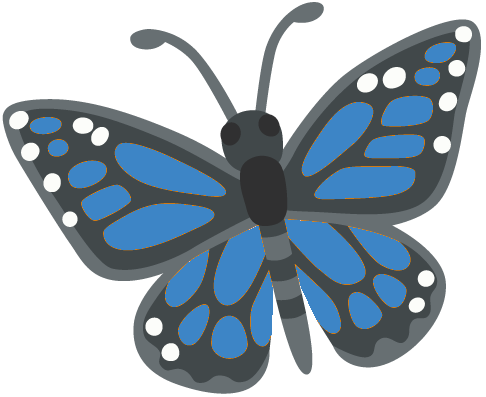}}{\textrm{C}}\xspace}
\newcommand{\butterflyred}{\scalerel*{\includegraphics{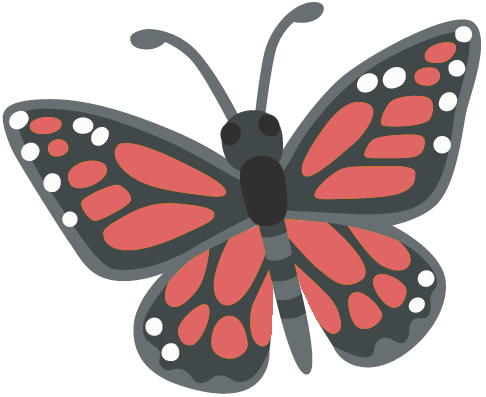}}{\textrm{C}}\xspace}
\newcommand{\butterflypurple}{\scalerel*{\includegraphics{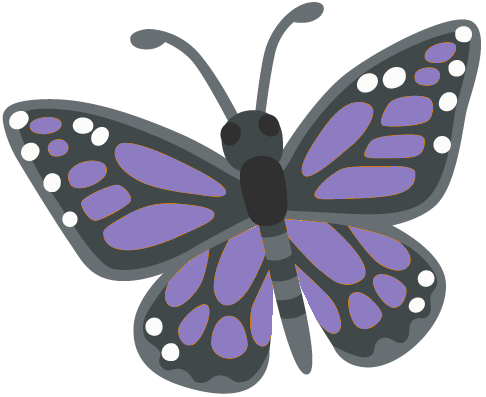}}{\textrm{C}}\xspace}
\newcommand{\butterflyyellow}{\scalerel*{\includegraphics{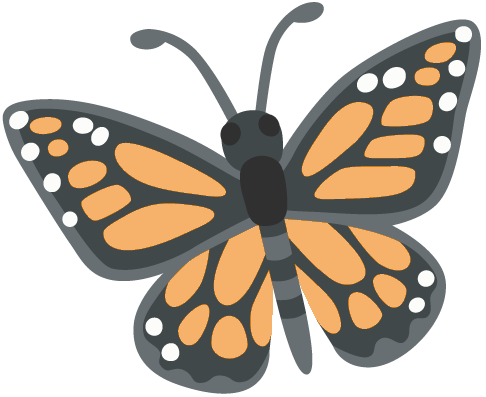}}{\textrm{C}}\xspace}
\newcommand{\butterflyturq}{\scalerel*{\includegraphics{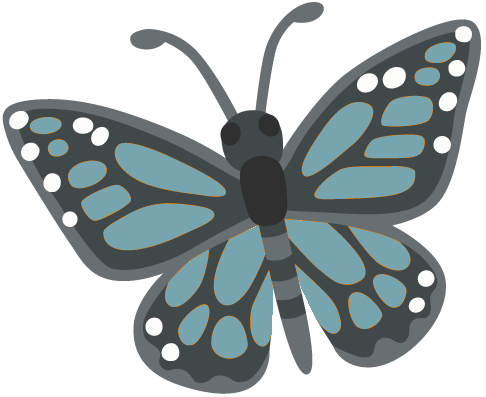}}{\textrm{C}}\xspace}

\newcommand{\cmark}{\ding{51}}
\newcommand{\xmark}{\ding{55}}

\definecolor{codegreen}{rgb}{0,0.6,0}
\lstdefinestyle{mystyle}{  
    commentstyle=\color{codegreen},
    keywordstyle=\color{blue},
    basicstyle=\ttfamily\small,
    breakatwhitespace=false,        
    breaklines=true,                 
    captionpos=b,                    
    keepspaces=true,                 
    showspaces=false,                
    showstringspaces=false,
    showtabs=false,                  
    tabsize=2
}
\definecolor{linkcolor}{HTML}{9c43fc}

\title{\textit{Top Leaderboard Ranking = Top Coding Proficiency, Always?}
\\
\butterfly{ }\dataset: Evolving Coding Benchmarks via \llm}

\newcommand{\seperate}{{\ \ \ \ \ \ \ \ \ \ \ \ \ \ \ \ \ \ \ \ \ \ \ \ \ \ \ \ \ \ \ \ \ \ \ \ }}
\newcommand{\sseparate}{{\ \ }}

\author{Chunqiu Steven Xia\textsuperscript{$*$} 
\sseparate
Yinlin Deng\thanks{Contributed equally with author ordering decided by \href{https://senseis.xmp.net/?Nigiri}{\textit{Nigiri}}.}
\seperate
Lingming Zhang
\\[\bigskipamount]
University of Illinois Urbana-Champaign{ }\uiuc{}
\\[\bigskipamount]
\texttt{\{chunqiu2, yinlind2, lingming\}@illinois.edu}
}

\colmfinalcopy %
\begin{document}

\maketitle

\begin{abstract}
\llm{s} have become the go-to choice for code generation tasks, with an exponential increase in the training, development, and usage of \llm{s} specifically for code generation.
To evaluate the ability of \llm{s} on code, both academic and industry practitioners rely on popular handcrafted benchmarks.
However, prior benchmarks contain only a very limited set of problems, both in quantity and variety.
Further, due to popularity and age, {many benchmarks} are prone to data leakage where example solutions can be readily found on the web and thus potentially in training data. 
Such limitations inevitably lead us to inquire:
\emph{Is the leaderboard performance on existing benchmarks reliable and comprehensive enough to measure the program synthesis ability of \llm{s}?}
To address this, we introduce \butterfly{ }\dataset -- a program synthesis benchmark suite created by \textit{evolving} existing benchmarks into different targeted domains for a comprehensive evaluation of \llm coding abilities. 
Our study on \textbf{\totalmodels} \llm{s} shows that compared to the high performance obtained on standard benchmarks like \humaneval, there is a significant drop in performance (on average \totaldecrease) when using \tech.
Additionally, the decrease in performance can range from \mindecrease to \maxdecrease, leading to drastic ranking changes amongst \llm{s} and showing potential overfitting of existing benchmarks.
Furthermore, we showcase various insights, including the brittleness of instruction-following models when encountering rewording or subtle changes as well as the importance of learning problem composition and decomposition. 
\tech not only provides comprehensive benchmarks, but can be used to further evolve arbitrary problems to keep up with advances and the ever-changing landscape of \llm{s} for code.
We have open-sourced our benchmarks, tools, and complete \llm generations at \textcolor{linkcolor}{\url{https://github.com/evo-eval/evoeval}}
\end{abstract}

\section{Introduction}

Program synthesis~\cite{PGL-010} is widely regarded as the \textit{holy-grail} in the field of computer science. 
Recently, \llmfull{s} (\llm{s}) have become the default choice for program synthesis due to its code reasoning capabilities acquired through training on large amounts of open-source code repositories. 
Popular \llm{s} like \gptfour~\cite{gptfour}, \claude{-3}~\cite{claudethree}, and \gemini~\cite{gemini} have shown tremendous success in aiding developers on a wide-range of coding tasks such as code completion~\cite{codex}, repair~\cite{codexrepair}, and test generation~\cite{titanfuzz}.  
Furthermore, researchers and industry practitioners have designed code \llm{s} (\eg \deepseek{ }Coder~\cite{deepseek}, \codellama~\cite{codellama}, and \starcoder~\cite{starcoder}) using a variety of training methods designed specifically for the code domain to improve \llm code understanding.

In order to evaluate the coding abilities of \llm{s}, benchmarks{ like} \humaneval~\cite{codex} and \mbpp~\cite{austin2021program} have been handcrafted to evaluate the program synthesis task of turning natural language descriptions (\eg docstrings) into code snippets. 
These code benchmarks measure functional correctness by evaluating \llm-generated solutions against a set of limited predefined tests.
Recent work~\cite{evalplus} has further improved these benchmarks with augmented tests to rigorously evaluate the functional correctness of \llm generated code.
However, apart from test inadequacy, existing popular code synthesis benchmarks have the following limitations: 

\begin{itemize}[noitemsep, leftmargin=*, topsep=0pt]
    \item \textbf{Limited amount and variety of problems.}
    Code benchmarks are mainly constructed by human annotators manually. 
    Due to the high manual effort required, they only contain a limited amount of problems. 
    For example, \humaneval~\cite{codex} only contains 164 handcrafted problems.
    Such a low amount of problems is not sufficient to fully measure the complete spectrum of program synthesis capability of state-of-the-art \llm{s}. 
    Additionally, these code benchmarks include mostly self-contained coding problems that lack variety in both problem types and domains, where the final evaluation output only shows the percentage of problems solved.
    While they provide a baseline overview of the coding abilities, 
    \llm builders and users cannot gain deeper insights to exactly what problem types or coding scenarios the particular \llm may excel or struggle in. 
    \item \textbf{Prone to data leakage and training dataset composition.} Popular benchmarks like \humaneval and \mbpp were released almost 4 years ago, with example solutions available in third-party open-source repositories. 
    While recent \llm{s} have been taking turns climbing the leaderboard by achieving higher \passat{1} scores (often with less than 1 percent difference between the next best model), just how much of that is attributed to having leaked solutions as part of the training data?
    Furthermore, the problems within these benchmarks are often simple derivatives of common coding problems/concepts.  
    In fact, recent work~\cite{riddell2024quantifying} has shown that there are substantial overlap between benchmark solutions and open-source training corpuses.
    In addition, closed-source \llm{s} may even deliberately include benchmark groundtruths to artificially boost their leaderboard status~\cite{balloccu2024leak}.
    As such, it is unclear whether high scores achieved by \llm{s} are truly due to their learnt coding capability or instead obtained via memorizing benchmark solutions.  
\end{itemize}

As more \llm{s} are being constructed, trained, and used especially for code, the insufficient evaluation benchmarks raise the question of validity: 
\emph{Is leaderboard performance on existing benchmarks reliable and comprehensive enough to measure the program synthesis ability of \llm{s}?}

\begin{figure}
    \includegraphics[width=\columnwidth]{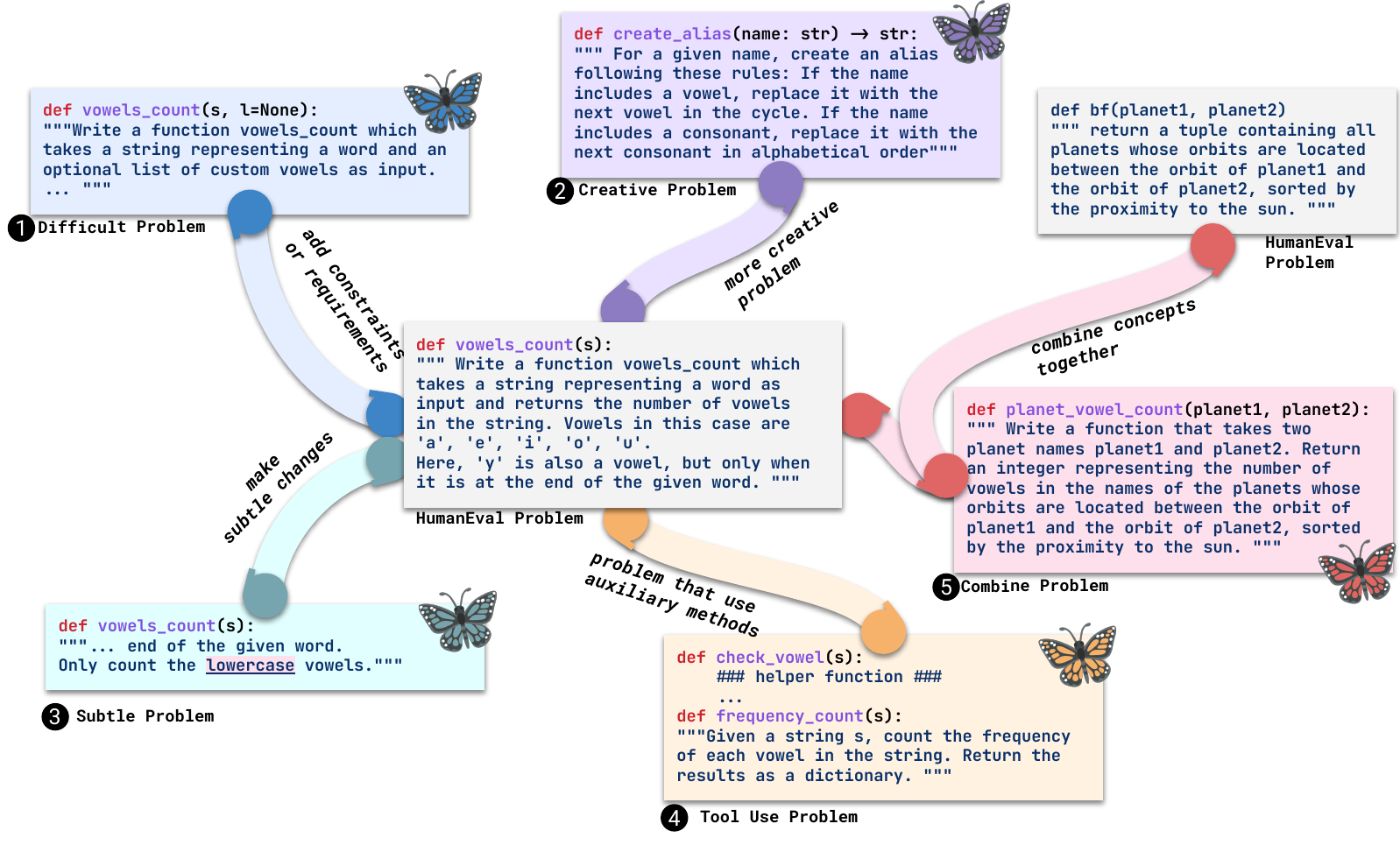}
    \centering
    \caption{Example problems generated in \tech through the use of targeted transformation prompts starting from a \humaneval problem.}
    \label{fig:quick_overview}
\end{figure}

\parabf{Our work.} To address the limitation of existing benchmarks,
we introduce \butterfly{ }\tech\footnote{coincidentally similar pronunciation with \textsc{EvilEval}} -- a set of program synthesis benchmarks created by \textit{evolving} existing problems. 
The key idea behind \tech is to use \llm{s} instead of humans to produce new code synthesis problems based on a variety of different instructions aimed at evolving or transforming the existing benchmark problems into targeted domains 
for more comprehensive evaluation.
Different from prior benchmark constructions that either obtain problems from open-source repositories or databases -- leading to data leakage or require manual construction of each problem -- resulting in high manual effort and limited diversity, \tech directly uses \llm{s} with targeted transformation prompts to synthesis new coding problems.
Specifically, we design 5 different targeted transformation prompts: \textit{Difficult, Creative, Subtle, Combine and Tool Use}.
We then prompt \gptfour{} to independently transform any existing problem in previous benchmarks into a new problem in the targeted domain. 

Figure~\ref{fig:quick_overview} shows a concrete example of \tech in action starting with an initial problem in \humaneval -- 
\CodeIn{vowel\_counts} to count the number of vowels in the string. 
\circled{1} We first observe the transformation to a more difficult problem by asking \gptfour to add additional constraints or requirements. 
This new problem contains a separate custom vowel list that makes the overall program logic more complex.
\circled{2} We can also transform to a more creative problem of \CodeIn{create\_alias} that still uses concepts like vowels and consonants but involves a much more creative and unusual problem description.
\circled{3} We can also make subtle changes to the problem where we only count the lowercase vowels to test if the \llm is simply memorizing the benchmark.
\circled{4} We can additionally combine concepts from multiple problems together. 
In the example, we use another problem \CodeIn{bf} to create a new problem that returns the vowels in each planet sorted based on the orbiting order. 
\circled{5} Furthermore, we can test the ability for \llm{s} to utilize auxiliary helper functions (common place in real-world code repositories) to solve more complex problems. 
Again we reuse the concepts of vowels from the initial problem, where the frequency of each vowel should be computed. However instead of directly solving the problem, the \llm can directly use the provided \CodeIn{check\_vowel} helper function to simplify the solution. 

Together, each of these transformed benchmarks are designed to introduce more difficult and complex problems as well as test different aspects of the \llm code understanding and synthesis ability. 
In \tech, we additionally use \gptfour{} to generate the groundtruth solution to each problem as well as rigorous test cases to ensure we can evaluate the functional correctness of \llm-synthesized code on \tech. 
Finally, we manually check each generated problem and corresponding groundtruth to ensure problem clarity and correctness. 
\tech serves as a way to further evolve existing benchmarks into more complex and well-suited problems for evaluation in order to keep up with the ever-growing \llm research. 

\parabf{Contribution.} Our work proposes to evolve existing problems for benchmark creation: 

\begin{itemize}[noitemsep, leftmargin=*, topsep=0pt]
    \item \textbf{Benchmark}: We present \tech -- a set of program synthesis benchmarks created by evolving existing popular \humaneval coding benchmark problems. 
    \tech includes 828 problems across 5 semantic-altering and 2 semantic-preserving benchmarks.
    Furthermore, \tech also includes additional benchmarks to study program synthesis concepts like problem composition and decomposition. %
    \tech is fully complete with groundtruth implementations and robust testcases to evaluate functional correctness.
    \item \textbf{Approach}: We propose a complete pipeline to directly synthesize new coding problems for benchmarking by evolving existing problems through the use of targeted transformation prompts. 
    Our pipeline aims to reduce manual checking effort using a self-consistency approach to automatically refine any problem inconsistencies and generate groundtruth as well as test cases.
    Our approach is general and can be used on other benchmark problems, adopted for transformation into additional domains or utilize different problem generation strategies~\cite{magicoder}.
    \item \textbf{Study}: We conduct a comprehensive study on \textbf{\totalmodels} different \llm{s} across all benchmarks in \tech.
    We found that compared to the high performance obtained on standard benchmarks like \humaneval, when evaluated on \tech, popular \llm{s} significantly drop in performance (on average \totaldecrease). 
    Additionally, this drop is not uniform across all \llm{s} and can range from \mindecrease to \maxdecrease, leading to drastic ranking changes amongst top performing models.
    We further demonstrate that certain \llm{s} cannot keep up their high performance obtained in \humaneval when evaluated on more challenging or problems in different domains, highlighting the possibilities of overfitting to existing benchmarks.
    Moreover, we observe that while instruction-following \llm{s} perform well in solving self-contained problems, they struggle with the tool using aspect of utilizing already provided auxiliary functions. 
    Furthermore, they are particularly sensitive to the problem description where rephrasing or subtle changes to the problem docstring leads to degradation in output solutions compared to their base non-instruction-following counterparts.  
    Additionally, we demonstrate that current state-of-the-art \llm{s} fail to effectively compose multiple general coding concepts to solve more complex variants, or address subproblems decomposed from previously solved difficult problem.
\end{itemize}

\section{Approach}
\label{sec:approach}

\begin{figure}
    \includegraphics[width=0.7\columnwidth]{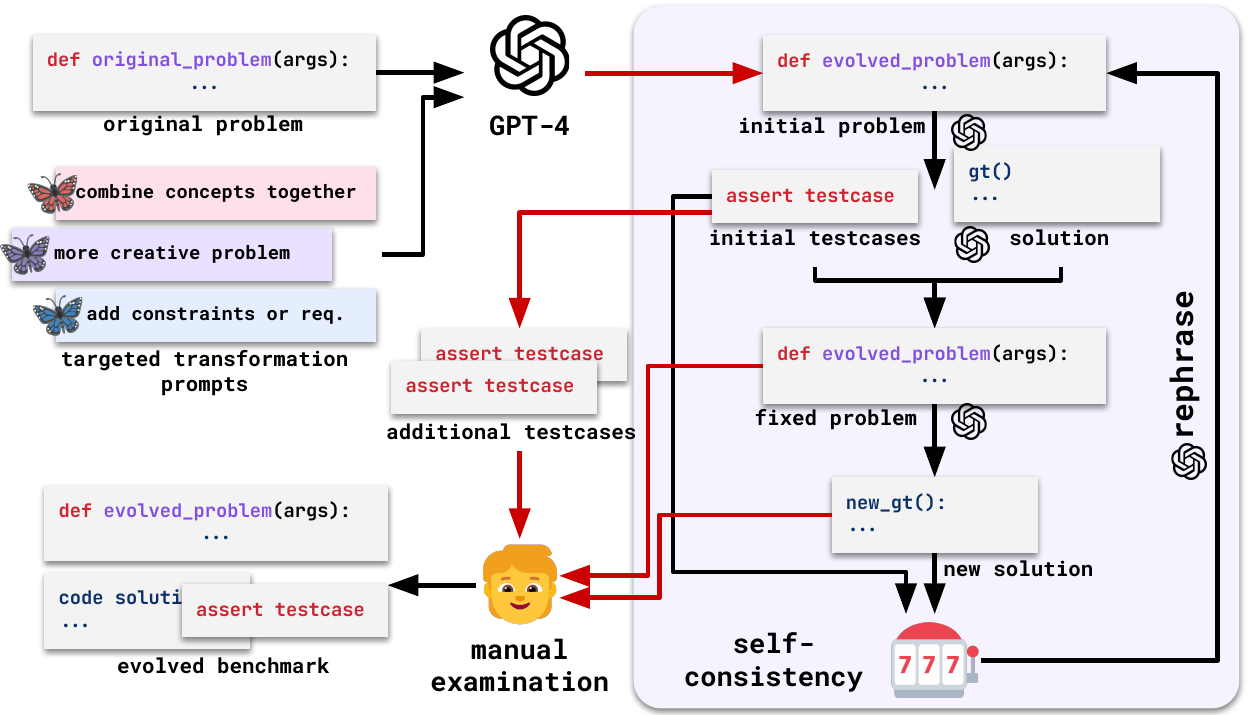}
    \centering
    \caption{Overview of \tech evolving problem generation pipeline.}
    \label{fig:overview}
\end{figure}

Figure~\ref{fig:overview} shows the overview of the benchmark creation pipeline for \tech.
We start by taking the original problem and apply a chosen targeted transformation prompt aimed at prompting \gptfour to produce a new code synthesis problem along the targeted domain.  
Using this initial transformed problem, we enter our refinement pipeline to fix any ambiguities or inconsistencies in the problem description, as well as generating the test cases and groundtruth solution for functional evaluation. 
Finally, to ensure correctness, we manually examine each produced problem along with the groundtruth and make corresponding changes to produce the final evolved benchmarks.  

\parabf{Targeted problem transformation.} 
\tech uses zero-shot prompting to evolve an existing coding benchmark to produce new and diverse problems. 
Each transformation prompt, as shown in the examples in Figure~\ref{fig:quick_overview}, aims to transform the existing problem in a specific manner.
In particular, we define two different types of transformation prompts: 1) \textbf{semantic-altering} -- change the semantic meaning of the original problem and 2) \textbf{semantic-preserving} -- modify the problem description while keeping the semantic meaning the same. 
While Figure~\ref{fig:quick_overview} shows only semantic-altering transformation prompts to produce new problems, we can also produce semantic-preserving problems to test additional aspect of the \llm coding abilities. 

\parabf{Problem refinement \& groundtruth Generation.} 
The initial evolved problem produced by \gptfour may include small inconsistencies such as contradicting sentences or incorrect I/O examples in the docstring.
For coding benchmarks, such inconsistencies are especially damaging as it can detract from the problem specification, leading to inaccurate evaluation of \llm coding capabilities.
As such, we introduce a refinement pipeline to iteratively rephrase and refine problem as needed. 
In addition, during this process, we also use \gptfour to produce the necessary groundtruth implementation of the function as well as example test cases to be used for evaluation. 

We first directly use \gptfour to obtain a possible solution for the initial problem. 
Additionally, we also prompt \gptfour to extract (if available in the initial problem docstring) or produce the test inputs for the transformed problem.
We then evaluate the test inputs on the solution to derive the corresponding expected test outputs.
Next, using these test inputs/outputs, we instruct \gptfour to add or fix the example test cases in the docstring, providing further demonstrations of the task.

Using this refined problem, we again generate a solution. 
We then leverage self-consistency~\cite{wang2022self} to check if the new solution on the test inputs produce the same outputs as the previous solution. 
The intuition is that since both solutions are generated by \gptfour and the refined problem should only include minimal changes (\eg adding new testcase examples), the solution output should then be the same in the absence of any potential inconsistencies or ambiguity in problem description. 
As such, if we observe differences between the two solution outputs, we ask \gptfour to further rephrase and fix any inconsistencies in the original problem and repeat the process.
On the other hand, if both solutions agree on outputs, we terminate the problem refinement stage and return the trio comprising of the new problem description, the solution as the groundtruth and the test cases for functional evaluation. 

\parabf{Manual examination \& test augmentation.} 
For each transformed problem, we carefully examine and adjust any final faults to ensure each problem and groundtruth is correctly specified and implemented. 
Additionally, using the initial set of test cases from the refinement stage, we further generate additional tests following the \llm-based test augmentation technique in \evalplus~\cite{evalplus}.
Finally, we produce \tech, a comprehensive code synthesis benchmark suite, which through the use of evolving transformations can generate diverse coding problems to evaluate \llm coding capability across various problem domains. 

\section{\tech Dataset Overview}
\label{sec:dataset}

\begin{table}
\setlength{\tabcolsep}{2pt}
  \centering
  \scriptsize
  \caption{\tech and \humaneval benchmark statistics. Note: the number in bracket shows the number of testcases in the augmented \humaneval{+} benchmarks, in \tech, they are directly reused in \eesubtle, \eeverbose and \eeconcise due the similarity. }\label{tab:dataset}
\begin{tabular}{l c ccccc cc}
\toprule
& original & \multicolumn{5}{c}{semantic-altering} & \multicolumn{2}{c}{semantic-preserving}  \\
\cmidrule(lr){2-2}
\cmidrule(lr){3-7}
\cmidrule(lr){8-9}
 & \humaneval & \eedifficult & \eecreative & \eesubtle & \eecombine & \eetoolusing & \eeverbose & \eeconcise \\
\midrule
\# problems & 164 & 100 & 100 & 100 & 100 & 100 & 164 & 164 \\
Avg. problem len. & 450.6 & 749.4 & 982.1 & 406.8 & 860.4 & 1224.6 & 450.6 & 450.6\\
Avg. \# test cases & 9.6 (764.1) & 49.8 & 43.1 & 10.3 (745.4) & 51.8 & 51.3 & 9.6 (764.1) & 9.6 (764.1) \\
\bottomrule
\end{tabular}
\end{table}

\begin{figure*}
    \centering
    \begin{subfigure}[b]{0.32\textwidth}
    \centering
    \includegraphics[width=\textwidth]{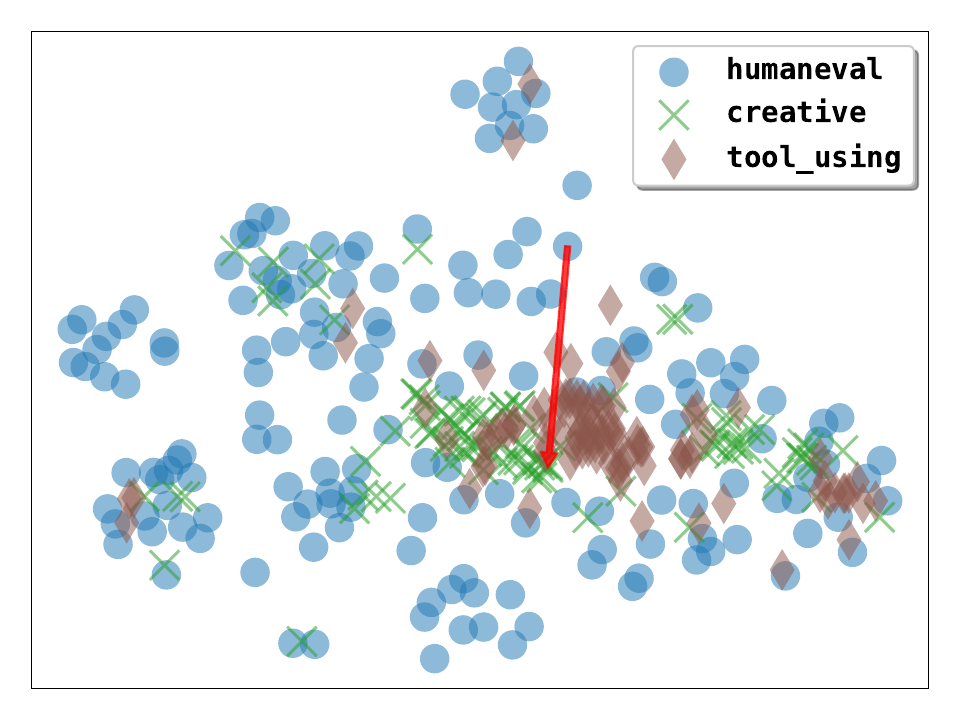}
    \caption{\eecreative \& \eetoolusing}
    \label{fig:creative_tool_using}
\end{subfigure}
\begin{subfigure}[b]{0.32\textwidth}
    \centering
    \includegraphics[width=\textwidth]{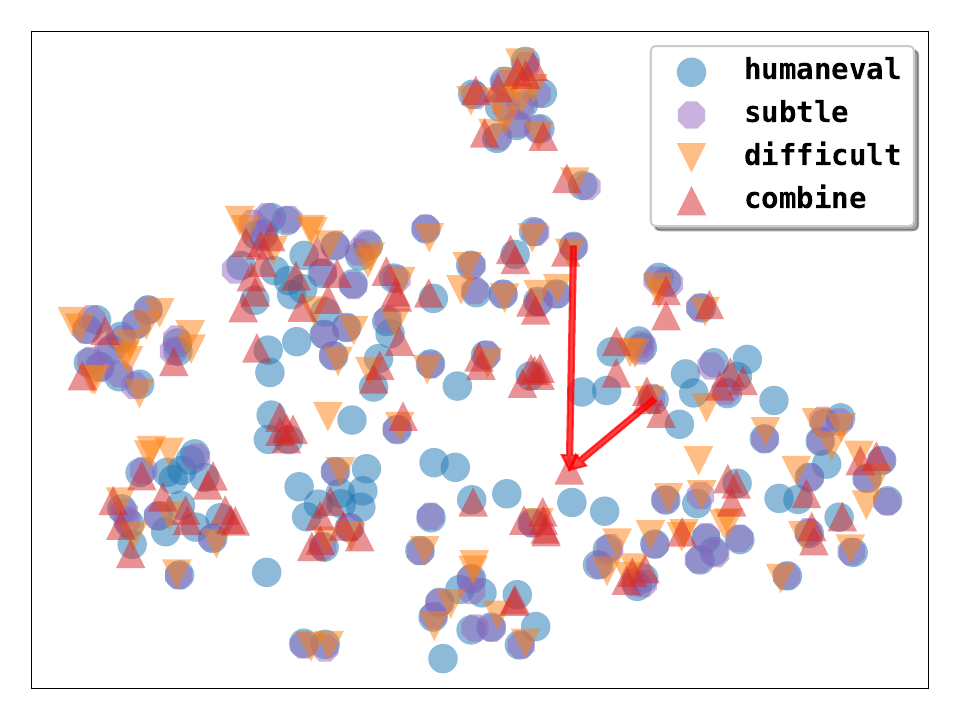}
    \caption{\eesubtle,\eedifficult,\eecombine}
    \label{fig:subtle_difficult_combine}
\end{subfigure}
\begin{subfigure}[b]{0.32\textwidth}
    \centering
    \includegraphics[width=\textwidth]{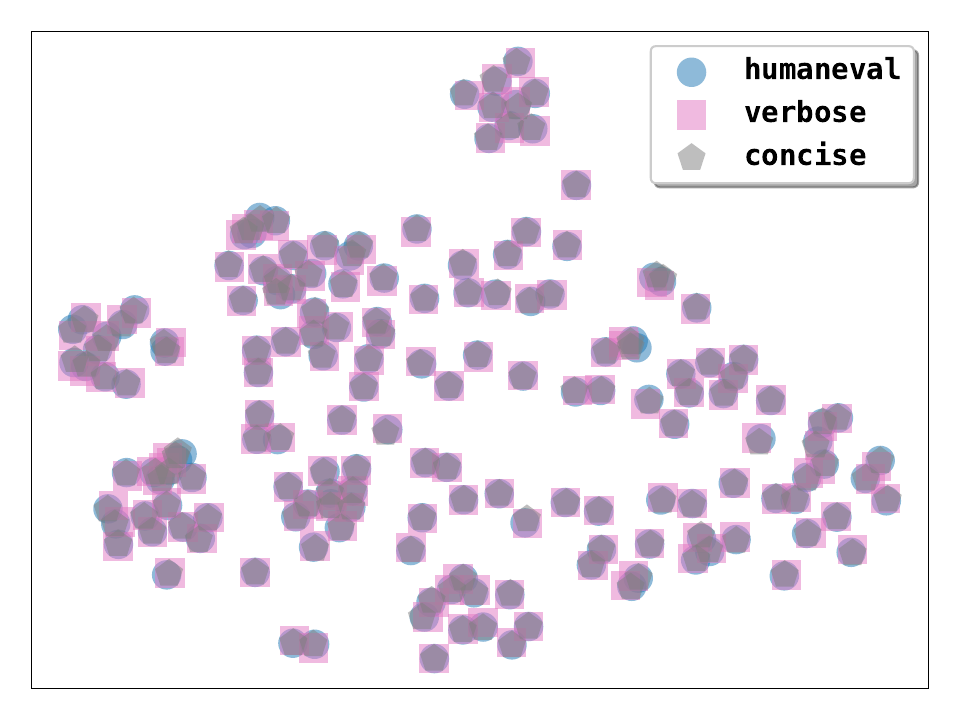}
    \caption{\eeverbose \& \eeconcise}
\end{subfigure}
\caption{2 dimensional t-SNE visualization of \tech benchmarks. }
\label{fig:tsne}
\end{figure*}

We use the problems in \humaneval as seeds to produce \tech. 
Problems in \tech consist mainly of self-contained functions, 
except for \eetoolusing that includes helper functions specifically designed to test the tool using capability of \llm{s}. 
Each problem uses a docstring to illustrate the problem specification, along with test cases and groundtruth to evaluate the functional correctness. 
Table~\ref{tab:dataset} shows the statistics of the benchmarks in \tech. 
In total, \tech includes 828 problems across 7 different datasets (5 semantic-altering and 2 semantic-preserving): 

\begin{itemize}[noitemsep, leftmargin=*, topsep=0pt]
    \item \textbf{\eedifficult}: Introduce complexity by adding additional constraints and requirements, replace commonly used requirements to less common ones, or add additional reasoning steps to the original problem. 
    \item \textbf{\eecreative}: Generate a more creative problem compared to the original through the use of stories or uncommon narratives. 
    \item \textbf{\eesubtle}: Make a subtle and minor change to the original problem such as inverting or replacing a requirement. 
    \item \textbf{\eecombine}: Combine two different problems by integrating the concepts from both problems. In order to select problems that make sense to combine, we apply a simple heuristic to combine only problems of the same type together categorized based on the type of input arguments in the original problem. 
    \item \textbf{\eetoolusing}: Produce a new problem containing a main problem and one or more helpers functions which can be used to solve it. 
    Each helper function is fully implemented and provides hints or useful functionality for solving the main problem. The main problem does not explicitly reference individual helper functions, and we do not require the model to use the provided helpers. %
    \item \textbf{\eeverbose}: Reword the original docstring to be more verbose. These verbose docstrings can use more descriptive language to illustrate the problem, include detailed explanation of the example output, and provide additional hints. 
    \item \textbf{\eeconcise}: Reword the original docstring to be more concise by removing unnecessary details and using concise language. 
    Furthermore, simple examples that are not required to demonstrate edge cases may be removed.
\end{itemize}

For each of the semantic-altering benchmarks, we generate 100 problems each using different seed problems from \humaneval. 
For semantic-preserving benchmarks, we generate using all 164 problems in \humaneval as it requires less validation since we can reuse the original groundtruths.
As shown in Table~\ref{tab:dataset}, compared to \humaneval, \tech contains longer coding questions with longer average problem length.
Furthermore, \tech also{ uses} more test cases to perform robust evaluation compared to base \humaneval.

Figure~\ref{fig:tsne} shows the embedding visualization using t-SNE~\cite{hinton2002stochastic}\footnote{perplexity=50 and iter=1000 using \CodeIn{text-embedding-3-large} model from OpenAI}
by projecting high-dimension representation of the problems docstrings in both \tech and \humaneval into the 2D plane.
First, we see that \eecreative and \eetoolusing drastically change the embedding distribution compared to the original dataset. 
The arrow in Figure~\ref{fig:creative_tool_using} shows one example of the shift in distribution from the original problem to a creative one.
Next, we see that \eesubtle, \eedifficult and \eecombine largely retain the same distribution as the original problems. 
This is due to the high parity across these problem description{s} where \eesubtle only applies subtle changes and \eedifficult adds additional complex constraints while keeping the main problem descriptions largely the same. 
Specifically, for \eecombine, we can see from an example arrow in Figure~\ref{fig:subtle_difficult_combine}, the new combined problem shifts the embedding for both of the original problems.
Finally, we observe that for \eeverbose and \eeconcise, the embeddings almost perfectly match the original problem, reflecting their semantic-preserving nature. 
In Appendix~\ref{sec:exampleprob}, we present example problems for each benchmark in \tech.

\section{Methodology}
\label{sec:method}

\parabf{Setup.} 
Each \llm generated sample is executed against the test cases in \tech and evaluated using differential testing~\cite{mckeeman1998differential} -- comparing against the groundtruth results to measure functional correctness.
We report the functional correctness by using the popular \passat{k} metric. 
We focus on greedy decoding (\ie producing a deterministic sample per each problem with temperature = 0). 
We denote this as \passat{1}. 

\parabf{Models.} We evaluate \textbf{\totalmodels} popular state-of-the-art \llm{s}, including both proprietary and open-source models {on \tech.}
We evaluate not only the popular general-purpose \llm{s} but also include recent code-based \llm{s} for comprehensive evaluation.
Further, we classify the \llm{s} as either base or instruction-following and focus our analysis on discussing the effect of model variants have on \tech performance. 

\parabf{Input format.} To produce the code solution using each \llm, we provide a specific input prompt: For base \llm{s} (\ie not instruction-tuned variants), we simply use only the function header with the docstring and let the \llm autocomplete the solution. 
For instruction-following \llm{s}, we follow the model-makers' guide on the exact instruction and format to use and ask the \llm to generate a complete solution for the problem.

\section{Evaluation}

\newcommand{\grow}[1]{ \textcolor{red}{\uparrow{#1}} }
\newcommand{\drop}[1]{ \textcolor{jwgreen}{\downarrow{#1}} }
\newcommand{\diff}[2]{ $\sfrac{\grow{#1}}{\drop{#2}}$ }
\newcommand{\exact}[2]{$\sfrac{#1}{#2}$}
\newcommand{\saturate}[1]{\textbf{#1}}
\newcommand{\ntest}[1]{\cellcolor{gray!20}{#1}}

\begin{table}
\setlength{\tabcolsep}{1pt}
  \caption{\passat{1} and ranking results (* indicates tie) on the semantically-altering \tech and \humaneval benchmarks. Note: \protect\speech denotes instruction-following \llm{s}. %
  }
  \label{tab:main}
  \centering
    \scriptsize
\begin{tabular}{lr rrrrrrrr rr rr rr}
\toprule & \multirow{2}*{Size}
         & \multicolumn{2}{c}{\humaneval}  
         & \multicolumn{2}{c}{\butterflyblue{}\eedifficult} 
         & \multicolumn{2}{c}{\butterflypurple{}\eecreative} 
         & \multicolumn{2}{c}{\butterflyturq{}\eesubtle}
         & \multicolumn{2}{c}{\butterflyred{}\eecombine} 
         & \multicolumn{2}{c}{\butterflyyellow{}\eetoolusing} 
         & \multicolumn{2}{c}{\butterflysmall{}\tech} \\
\cmidrule(lr){3-4}
\cmidrule(lr){5-6}
\cmidrule(lr){7-8}
\cmidrule(lr){9-10}
\cmidrule(lr){11-12}
\cmidrule(lr){13-14}
\cmidrule(lr){15-16}
         &  
         & \passat{1} & rank
         & \passat{1} & rank
         & \passat{1} & rank
         & \passat{1} & rank
         & \passat{1} & rank
         & \passat{1} & rank
         & \passat{1} & rank \\
\midrule

\openai{ }\gptfour{-Turbo}{\speech} & NA & \cellcolor[HTML]{f3f3ff}83.5 (80.5) & \textbf{\textcolor{red}{1}} & \cellcolor[HTML]{e1ecff}50.0 & \textbf{\textcolor{red}{*2}} & \cellcolor[HTML]{e9e1ff}61.0 & \textbf{\textcolor{red}{2}} & \cellcolor[HTML]{e1ffff}82.0 & \textbf{\textcolor{red}{1}} & \cellcolor[HTML]{ffe1ee}45.0 & \textbf{\textcolor{red}{2}} & \cellcolor[HTML]{fff2e1}69.0 & \textbf{\textcolor{red}{*1}} & 65.1 & \textbf{\textcolor{red}{2}}\\
\midrule
\openai{ }\gptfour{\speech} & NA & \cellcolor[HTML]{f3f3ff}82.3 (76.2) & \textbf{\textcolor{red}{*2}} & \cellcolor[HTML]{e1ecff}52.0 & \textbf{\textcolor{red}{1}} & \cellcolor[HTML]{e9e1ff}66.0 & \textbf{\textcolor{red}{1}} & \cellcolor[HTML]{e1ffff}76.0 & \textbf{\textcolor{red}{3}} & \cellcolor[HTML]{ffe1ee}53.0 & \textbf{\textcolor{red}{1}} & \cellcolor[HTML]{fff2e1}68.0 & \textbf{\textcolor{red}{3}} & 66.2 & \textbf{\textcolor{red}{1}}\\
\midrule
\openai{ }\chatgpt{\speech} & NA & \cellcolor[HTML]{f3f3ff}76.8 (69.5) & *5 & \cellcolor[HTML]{e1ecff}33.0 & *13 & \cellcolor[HTML]{e9e1ff}42.0 & *7 & \cellcolor[HTML]{e1ffff}70.0 & 4 & \cellcolor[HTML]{ffe1ee}33.0 & 4 & \cellcolor[HTML]{fff2e1}64.0 & *6 & 53.1 & 6\\
\midrule
\anthropic{ }\claude{-3}{\speech} & NA & \cellcolor[HTML]{f3f3ff}82.3 (75.0) & \textbf{\textcolor{red}{*2}} & \cellcolor[HTML]{e1ecff}50.0 & \textbf{\textcolor{red}{*2}} & \cellcolor[HTML]{e9e1ff}53.0 & \textbf{\textcolor{red}{3}} & \cellcolor[HTML]{e1ffff}81.0 & \textbf{\textcolor{red}{2}} & \cellcolor[HTML]{ffe1ee}42.0 & \textbf{\textcolor{red}{3}} & \cellcolor[HTML]{fff2e1}69.0 & \textbf{\textcolor{red}{*1}} & 62.9 & \textbf{\textcolor{red}{3}}\\
\midrule
\anthropic{ }\claude{-3}{-haiku}{\speech} & NA & \cellcolor[HTML]{f3f3ff}74.4 (66.5) & *8 & \cellcolor[HTML]{e1ecff}40.0 & *6 & \cellcolor[HTML]{e9e1ff}47.0 & *5 & \cellcolor[HTML]{e1ffff}65.0 & *10 & \cellcolor[HTML]{ffe1ee}25.0 & *6 & \cellcolor[HTML]{fff2e1}61.0 & *10 & 52.1 & 7\\
\midrule
\anthropic{ }\claude{-2}{\speech} & NA & \cellcolor[HTML]{f3f3ff}66.5 (62.2) & *18 & \cellcolor[HTML]{e1ecff}29.0 & 17 & \cellcolor[HTML]{e9e1ff}42.0 & *7 & \cellcolor[HTML]{e1ffff}64.0 & *13 & \cellcolor[HTML]{ffe1ee}19.0 & 14 & \cellcolor[HTML]{fff2e1}57.0 & *16 & 46.2 & 15\\
\midrule
\google{ }\gemini{\speech} & NA & \cellcolor[HTML]{f3f3ff}62.2 (56.7) & 21 & \cellcolor[HTML]{e1ecff}37.0 & *10 & \cellcolor[HTML]{e9e1ff}40.0 & 12 & \cellcolor[HTML]{e1ffff}53.0 & *21 & \cellcolor[HTML]{ffe1ee}23.0 & *9 & \cellcolor[HTML]{fff2e1}57.0 & *16 & 45.4 & 17\\
\midrule
\google{ }\palm{\speech} & NA & \cellcolor[HTML]{f3f3ff}40.2 (36.6) & 38 & \cellcolor[HTML]{e1ecff}18.0 & *32 & \cellcolor[HTML]{e9e1ff}22.0 & 33 & \cellcolor[HTML]{e1ffff}36.0 & *42 & \cellcolor[HTML]{ffe1ee}3.0 & *39 & \cellcolor[HTML]{fff2e1}46.0 & *29 & 27.5 & 37\\
\midrule
\multirow{3}*{\deepseeklogo{ }\deepseekinstruct{\speech}} & 33b & \cellcolor[HTML]{f3f3ff}78.0 (73.2) & 4 & \cellcolor[HTML]{e1ecff}47.0 & 5 & \cellcolor[HTML]{e9e1ff}47.0 & *5 & \cellcolor[HTML]{e1ffff}67.0 & *5 & \cellcolor[HTML]{ffe1ee}31.0 & 5 & \cellcolor[HTML]{fff2e1}66.0 & 4 & 56.0 & 4\\
 & 6.7b & \cellcolor[HTML]{f3f3ff}74.4 (69.5) & *8 & \cellcolor[HTML]{e1ecff}40.0 & *6 & \cellcolor[HTML]{e9e1ff}37.0 & *13 & \cellcolor[HTML]{e1ffff}61.0 & *17 & \cellcolor[HTML]{ffe1ee}18.0 & *15 & \cellcolor[HTML]{fff2e1}51.0 & 24 & 46.9 & 14\\
 & 1.3b & \cellcolor[HTML]{f3f3ff}63.4 (60.4) & 20 & \cellcolor[HTML]{e1ecff}20.0 & *30 & \cellcolor[HTML]{e9e1ff}25.0 & *25 & \cellcolor[HTML]{e1ffff}53.0 & *21 & \cellcolor[HTML]{ffe1ee}9.0 & *28 & \cellcolor[HTML]{fff2e1}39.0 & *41 & 34.9 & 24\\
\midrule
\multirow{3}*{\deepseeklogo{ }\deepseek} & 33b & \cellcolor[HTML]{f3f3ff}50.6 (42.7) & 26 & \cellcolor[HTML]{e1ecff}26.0 & 20 & \cellcolor[HTML]{e9e1ff}23.0 & *30 & \cellcolor[HTML]{e1ffff}47.0 & *26 & \cellcolor[HTML]{ffe1ee}11.0 & *25 & \cellcolor[HTML]{fff2e1}63.0 & *8 & 36.8 & 23\\
 & 6.7b & \cellcolor[HTML]{f3f3ff}45.1 (38.4) & *31 & \cellcolor[HTML]{e1ecff}21.0 & *26 & \cellcolor[HTML]{e9e1ff}24.0 & *27 & \cellcolor[HTML]{e1ffff}47.0 & *26 & \cellcolor[HTML]{ffe1ee}5.0 & *35 & \cellcolor[HTML]{fff2e1}55.0 & *19 & 32.9 & 29\\
 & 1.3b & \cellcolor[HTML]{f3f3ff}29.9 (26.2) & 45 & \cellcolor[HTML]{e1ecff}6.0 & *48 & \cellcolor[HTML]{e9e1ff}19.0 & *35 & \cellcolor[HTML]{e1ffff}27.0 & 49 & \cellcolor[HTML]{ffe1ee}0.0 & 51 & \cellcolor[HTML]{fff2e1}40.0 & 40 & 20.3 & 45\\
\midrule
\deepseeklogo{ }\deepseekvonefive{-Inst}{\speech} & 7b & \cellcolor[HTML]{f3f3ff}68.9 (63.4) & *15 & \cellcolor[HTML]{e1ecff}37.0 & *10 & \cellcolor[HTML]{e9e1ff}37.0 & *13 & \cellcolor[HTML]{e1ffff}66.0 & *8 & \cellcolor[HTML]{ffe1ee}24.0 & 8 & \cellcolor[HTML]{fff2e1}60.0 & *12 & 48.8 & 10\\
\midrule
\deepseeklogo{ }\deepseekvonefive & 7b & \cellcolor[HTML]{f3f3ff}42.1 (34.8) & *35 & \cellcolor[HTML]{e1ecff}21.0 & *26 & \cellcolor[HTML]{e9e1ff}34.0 & *17 & \cellcolor[HTML]{e1ffff}43.0 & *31 & \cellcolor[HTML]{ffe1ee}4.0 & *37 & \cellcolor[HTML]{fff2e1}54.0 & *21 & 33.0 & 28\\
\midrule
\multirow{4}*{\meta{ }\codellamainstruct{\speech}} & 70b & \cellcolor[HTML]{f3f3ff}66.5 (59.8) & *18 & \cellcolor[HTML]{e1ecff}31.0 & 16 & \cellcolor[HTML]{e9e1ff}41.0 & *10 & \cellcolor[HTML]{e1ffff}65.0 & *10 & \cellcolor[HTML]{ffe1ee}18.0 & *15 & \cellcolor[HTML]{fff2e1}65.0 & 5 & 47.7 & 12\\
 & 34b & \cellcolor[HTML]{f3f3ff}51.8 (43.9) & 25 & \cellcolor[HTML]{e1ecff}22.0 & *24 & \cellcolor[HTML]{e9e1ff}27.0 & 23 & \cellcolor[HTML]{e1ffff}43.0 & *31 & \cellcolor[HTML]{ffe1ee}9.0 & *28 & \cellcolor[HTML]{fff2e1}47.0 & *27 & 33.3 & 27\\
 & 13b & \cellcolor[HTML]{f3f3ff}48.8 (42.7) & 29 & \cellcolor[HTML]{e1ecff}21.0 & *26 & \cellcolor[HTML]{e9e1ff}25.0 & *25 & \cellcolor[HTML]{e1ffff}46.0 & 29 & \cellcolor[HTML]{ffe1ee}8.0 & *31 & \cellcolor[HTML]{fff2e1}54.0 & *21 & 33.8 & 26\\
 & 7b & \cellcolor[HTML]{f3f3ff}43.3 (39.0) & 33 & \cellcolor[HTML]{e1ecff}14.0 & 38 & \cellcolor[HTML]{e9e1ff}18.0 & *37 & \cellcolor[HTML]{e1ffff}40.0 & *37 & \cellcolor[HTML]{ffe1ee}8.0 & *31 & \cellcolor[HTML]{fff2e1}44.0 & *33 & 27.9 & 36\\
\midrule
\multirow{4}*{\meta{ }\codellama} & 70b & \cellcolor[HTML]{f3f3ff}60.4 (52.4) & 23 & \cellcolor[HTML]{e1ecff}25.0 & 21 & \cellcolor[HTML]{e9e1ff}29.0 & *20 & \cellcolor[HTML]{e1ffff}49.0 & *23 & \cellcolor[HTML]{ffe1ee}14.0 & *21 & \cellcolor[HTML]{fff2e1}63.0 & *8 & 40.1 & 22\\
 & 34b & \cellcolor[HTML]{f3f3ff}52.4 (43.3) & 24 & \cellcolor[HTML]{e1ecff}15.0 & 37 & \cellcolor[HTML]{e9e1ff}24.0 & *27 & \cellcolor[HTML]{e1ffff}47.0 & *26 & \cellcolor[HTML]{ffe1ee}11.0 & *25 & \cellcolor[HTML]{fff2e1}44.0 & *33 & 32.2 & 30\\
 & 13b & \cellcolor[HTML]{f3f3ff}42.7 (36.6) & 34 & \cellcolor[HTML]{e1ecff}18.0 & *32 & \cellcolor[HTML]{e9e1ff}24.0 & *27 & \cellcolor[HTML]{e1ffff}38.0 & *39 & \cellcolor[HTML]{ffe1ee}6.0 & 34 & \cellcolor[HTML]{fff2e1}48.0 & *25 & 29.4 & 33\\
 & 7b & \cellcolor[HTML]{f3f3ff}39.6 (36.6) & 39 & \cellcolor[HTML]{e1ecff}10.0 & *42 & \cellcolor[HTML]{e9e1ff}15.0 & 41 & \cellcolor[HTML]{e1ffff}42.0 & 34 & \cellcolor[HTML]{ffe1ee}3.0 & *39 & \cellcolor[HTML]{fff2e1}44.0 & *33 & 25.6 & 38\\
\midrule
\wizardcoder{\speech} & 34b & \cellcolor[HTML]{f3f3ff}61.6 (54.3) & 22 & \cellcolor[HTML]{e1ecff}24.0 & 22 & \cellcolor[HTML]{e9e1ff}32.0 & 19 & \cellcolor[HTML]{e1ffff}55.0 & 20 & \cellcolor[HTML]{ffe1ee}17.0 & *18 & \cellcolor[HTML]{fff2e1}55.0 & *19 & 40.8 & 20\\
\midrule
\wizardcoder{-1.1}{\speech} & 33b & \cellcolor[HTML]{f3f3ff}73.8 (69.5) & 10 & \cellcolor[HTML]{e1ecff}48.0 & 4 & \cellcolor[HTML]{e9e1ff}48.0 & 4 & \cellcolor[HTML]{e1ffff}66.0 & *8 & \cellcolor[HTML]{ffe1ee}20.0 & 13 & \cellcolor[HTML]{fff2e1}64.0 & *6 & 53.3 & 5\\
\midrule
\xwincoder{\speech} & 34b & \cellcolor[HTML]{f3f3ff}68.9 (62.2) & *15 & \cellcolor[HTML]{e1ecff}33.0 & *13 & \cellcolor[HTML]{e9e1ff}42.0 & *7 & \cellcolor[HTML]{e1ffff}67.0 & *5 & \cellcolor[HTML]{ffe1ee}15.0 & 20 & \cellcolor[HTML]{fff2e1}60.0 & *12 & 47.7 & 13\\
\midrule
\phindllamatwo & 34b & \cellcolor[HTML]{f3f3ff}70.7 (66.5) & 13 & \cellcolor[HTML]{e1ecff}22.0 & *24 & \cellcolor[HTML]{e9e1ff}35.0 & 16 & \cellcolor[HTML]{e1ffff}63.0 & 15 & \cellcolor[HTML]{ffe1ee}25.0 & *6 & \cellcolor[HTML]{fff2e1}58.0 & 15 & 45.6 & 16\\
\midrule
\codemillenials{\speech} & 34b & \cellcolor[HTML]{f3f3ff}73.2 (69.5) & 11 & \cellcolor[HTML]{e1ecff}35.0 & 12 & \cellcolor[HTML]{e9e1ff}41.0 & *10 & \cellcolor[HTML]{e1ffff}65.0 & *10 & \cellcolor[HTML]{ffe1ee}17.0 & *18 & \cellcolor[HTML]{fff2e1}56.0 & 18 & 47.9 & 11\\
\midrule
\speechlesscodellama{\speech} & 34b & \cellcolor[HTML]{f3f3ff}75.0 (69.5) & 7 & \cellcolor[HTML]{e1ecff}38.0 & 9 & \cellcolor[HTML]{e9e1ff}37.0 & *13 & \cellcolor[HTML]{e1ffff}64.0 & *13 & \cellcolor[HTML]{ffe1ee}23.0 & *9 & \cellcolor[HTML]{fff2e1}59.0 & 14 & 49.3 & 9\\
\midrule
\magicoder{-s-DS}{\speech} & 6.7b & \cellcolor[HTML]{f3f3ff}76.8 (70.7) & *5 & \cellcolor[HTML]{e1ecff}40.0 & *6 & \cellcolor[HTML]{e9e1ff}34.0 & *17 & \cellcolor[HTML]{e1ffff}67.0 & *5 & \cellcolor[HTML]{ffe1ee}21.0 & *11 & \cellcolor[HTML]{fff2e1}61.0 & *10 & 50.0 & 8\\
\midrule
\magicoder{-s-CL}{\speech} & 7b & \cellcolor[HTML]{f3f3ff}70.1 (65.9) & 14 & \cellcolor[HTML]{e1ecff}27.0 & 19 & \cellcolor[HTML]{e9e1ff}26.0 & 24 & \cellcolor[HTML]{e1ffff}58.0 & 19 & \cellcolor[HTML]{ffe1ee}11.0 & *25 & \cellcolor[HTML]{fff2e1}52.0 & 23 & 40.7 & 21\\
\midrule
\multirow{3}*{\starcodertwo} & 15b & \cellcolor[HTML]{f3f3ff}45.1 (36.0) & *31 & \cellcolor[HTML]{e1ecff}16.0 & *35 & \cellcolor[HTML]{e9e1ff}19.0 & *35 & \cellcolor[HTML]{e1ffff}41.0 & *35 & \cellcolor[HTML]{ffe1ee}5.0 & *35 & \cellcolor[HTML]{fff2e1}48.0 & *25 & 29.0 & 35\\
 & 7b & \cellcolor[HTML]{f3f3ff}34.8 (31.1) & *40 & \cellcolor[HTML]{e1ecff}12.0 & *39 & \cellcolor[HTML]{e9e1ff}17.0 & 39 & \cellcolor[HTML]{e1ffff}38.0 & *39 & \cellcolor[HTML]{ffe1ee}2.0 & *45 & \cellcolor[HTML]{fff2e1}46.0 & *29 & 25.0 & 40\\
 & 3b & \cellcolor[HTML]{f3f3ff}31.1 (26.2) & 44 & \cellcolor[HTML]{e1ecff}8.0 & *45 & \cellcolor[HTML]{e9e1ff}14.0 & *42 & \cellcolor[HTML]{e1ffff}31.0 & *44 & \cellcolor[HTML]{ffe1ee}2.0 & *45 & \cellcolor[HTML]{fff2e1}35.0 & 45 & 20.2 & 46\\
\midrule
\starcoder & 15b & \cellcolor[HTML]{f3f3ff}34.8 (30.5) & *40 & \cellcolor[HTML]{e1ecff}12.0 & *39 & \cellcolor[HTML]{e9e1ff}11.0 & 47 & \cellcolor[HTML]{e1ffff}37.0 & 41 & \cellcolor[HTML]{ffe1ee}2.0 & *45 & \cellcolor[HTML]{fff2e1}44.0 & *33 & 23.5 & 41\\
\midrule
\mistrallogo{ }\mixtralinstruct{\speech} & 8x7b & \cellcolor[HTML]{f3f3ff}42.1 (38.4) & *35 & \cellcolor[HTML]{e1ecff}21.0 & *26 & \cellcolor[HTML]{e9e1ff}18.0 & *37 & \cellcolor[HTML]{e1ffff}41.0 & *35 & \cellcolor[HTML]{ffe1ee}9.0 & *28 & \cellcolor[HTML]{fff2e1}45.0 & *31 & 29.3 & 34\\
\midrule
\mistrallogo{ }\mistralinstruct{-v02}{\speech} & 7b & \cellcolor[HTML]{f3f3ff}28.0 (23.2) & *48 & \cellcolor[HTML]{e1ecff}8.0 & *45 & \cellcolor[HTML]{e9e1ff}16.0 & 40 & \cellcolor[HTML]{e1ffff}25.0 & 50 & \cellcolor[HTML]{ffe1ee}3.0 & *39 & \cellcolor[HTML]{fff2e1}8.0 & 51 & 14.7 & 50\\
\midrule
\mistrallogo{ }\mistralinstruct{\speech} & 7b & \cellcolor[HTML]{f3f3ff}28.7 (24.4) & 47 & \cellcolor[HTML]{e1ecff}6.0 & *48 & \cellcolor[HTML]{e9e1ff}8.0 & 50 & \cellcolor[HTML]{e1ffff}31.0 & *44 & \cellcolor[HTML]{ffe1ee}3.0 & *39 & \cellcolor[HTML]{fff2e1}29.0 & 48 & 17.6 & 49\\
\midrule
\mistrallogo{ }\mistral & 7b & \cellcolor[HTML]{f3f3ff}28.0 (23.8) & *48 & \cellcolor[HTML]{e1ecff}8.0 & *45 & \cellcolor[HTML]{e9e1ff}14.0 & *42 & \cellcolor[HTML]{e1ffff}30.0 & 47 & \cellcolor[HTML]{ffe1ee}3.0 & *39 & \cellcolor[HTML]{fff2e1}38.0 & 43 & 20.2 & 47\\
\midrule
\openchat{\speech} & 7b & \cellcolor[HTML]{f3f3ff}71.3 (66.5) & 12 & \cellcolor[HTML]{e1ecff}33.0 & *13 & \cellcolor[HTML]{e9e1ff}29.0 & *20 & \cellcolor[HTML]{e1ffff}62.0 & 16 & \cellcolor[HTML]{ffe1ee}14.0 & *21 & \cellcolor[HTML]{fff2e1}43.0 & 38 & 42.1 & 18\\
\midrule
\stablecode & 3b & \cellcolor[HTML]{f3f3ff}29.3 (25.6) & 46 & \cellcolor[HTML]{e1ecff}10.0 & *42 & \cellcolor[HTML]{e9e1ff}10.0 & *48 & \cellcolor[HTML]{e1ffff}31.0 & *44 & \cellcolor[HTML]{ffe1ee}3.0 & *39 & \cellcolor[HTML]{fff2e1}41.0 & 39 & 20.7 & 43\\
\midrule
\google{ }\gemma{-Inst}{\speech} & 7b & \cellcolor[HTML]{f3f3ff}28.0 (23.2) & *48 & \cellcolor[HTML]{e1ecff}6.0 & *48 & \cellcolor[HTML]{e9e1ff}10.0 & *48 & \cellcolor[HTML]{e1ffff}29.0 & 48 & \cellcolor[HTML]{ffe1ee}2.0 & *45 & \cellcolor[HTML]{fff2e1}31.0 & 47 & 17.7 & 48\\
\midrule
\multirow{2}*{\google{ }\gemma} & 7b & \cellcolor[HTML]{f3f3ff}31.7 (25.0) & 43 & \cellcolor[HTML]{e1ecff}12.0 & *39 & \cellcolor[HTML]{e9e1ff}13.0 & *44 & \cellcolor[HTML]{e1ffff}40.0 & *37 & \cellcolor[HTML]{ffe1ee}2.0 & *45 & \cellcolor[HTML]{fff2e1}39.0 & *41 & 23.0 & 42\\
 & 2b & \cellcolor[HTML]{f3f3ff}22.0 (17.1) & 51 & \cellcolor[HTML]{e1ecff}2.0 & 51 & \cellcolor[HTML]{e9e1ff}6.0 & 51 & \cellcolor[HTML]{e1ffff}24.0 & 51 & \cellcolor[HTML]{ffe1ee}2.0 & *45 & \cellcolor[HTML]{fff2e1}21.0 & 50 & 12.8 & 51\\
\midrule
\microsoft{ }\phitwo & 2.7b & \cellcolor[HTML]{f3f3ff}50.0 (45.1) & *27 & \cellcolor[HTML]{e1ecff}18.0 & *32 & \cellcolor[HTML]{e9e1ff}23.0 & *30 & \cellcolor[HTML]{e1ffff}49.0 & *23 & \cellcolor[HTML]{ffe1ee}14.0 & *21 & \cellcolor[HTML]{fff2e1}37.0 & 44 & 31.8 & 32\\
\midrule
\multirow{3}*{\qwen{\speech}} & 72b & \cellcolor[HTML]{f3f3ff}67.1 (61.6) & 17 & \cellcolor[HTML]{e1ecff}28.0 & 18 & \cellcolor[HTML]{e9e1ff}28.0 & 22 & \cellcolor[HTML]{e1ffff}61.0 & *17 & \cellcolor[HTML]{ffe1ee}21.0 & *11 & \cellcolor[HTML]{fff2e1}47.0 & *27 & 42.0 & 19\\
 & 14b & \cellcolor[HTML]{f3f3ff}50.0 (45.7) & *27 & \cellcolor[HTML]{e1ecff}20.0 & *30 & \cellcolor[HTML]{e9e1ff}23.0 & *30 & \cellcolor[HTML]{e1ffff}48.0 & 25 & \cellcolor[HTML]{ffe1ee}18.0 & *15 & \cellcolor[HTML]{fff2e1}44.0 & *33 & 33.8 & 25\\
 & 7b & \cellcolor[HTML]{f3f3ff}42.1 (37.8) & *35 & \cellcolor[HTML]{e1ecff}16.0 & *35 & \cellcolor[HTML]{e9e1ff}13.0 & *44 & \cellcolor[HTML]{e1ffff}43.0 & *31 & \cellcolor[HTML]{ffe1ee}7.0 & 33 & \cellcolor[HTML]{fff2e1}32.0 & 46 & 25.5 & 39\\
\midrule
\multirow{2}*{\qwenb{\speech}} & 14b & \cellcolor[HTML]{f3f3ff}46.3 (43.9) & 30 & \cellcolor[HTML]{e1ecff}23.0 & 23 & \cellcolor[HTML]{e9e1ff}20.0 & 34 & \cellcolor[HTML]{e1ffff}45.0 & 30 & \cellcolor[HTML]{ffe1ee}13.0 & 24 & \cellcolor[HTML]{fff2e1}45.0 & *31 & 32.1 & 31\\
 & 7b & \cellcolor[HTML]{f3f3ff}34.1 (29.9) & 42 & \cellcolor[HTML]{e1ecff}9.0 & 44 & \cellcolor[HTML]{e9e1ff}12.0 & 46 & \cellcolor[HTML]{e1ffff}36.0 & *42 & \cellcolor[HTML]{ffe1ee}4.0 & *37 & \cellcolor[HTML]{fff2e1}28.0 & 49 & 20.5 & 44\\
\bottomrule
  \end{tabular}
\end{table}

\subsection{\llm Synthesis \& Evaluation on \tech}

\begin{figure}
    \includegraphics[width=\columnwidth]{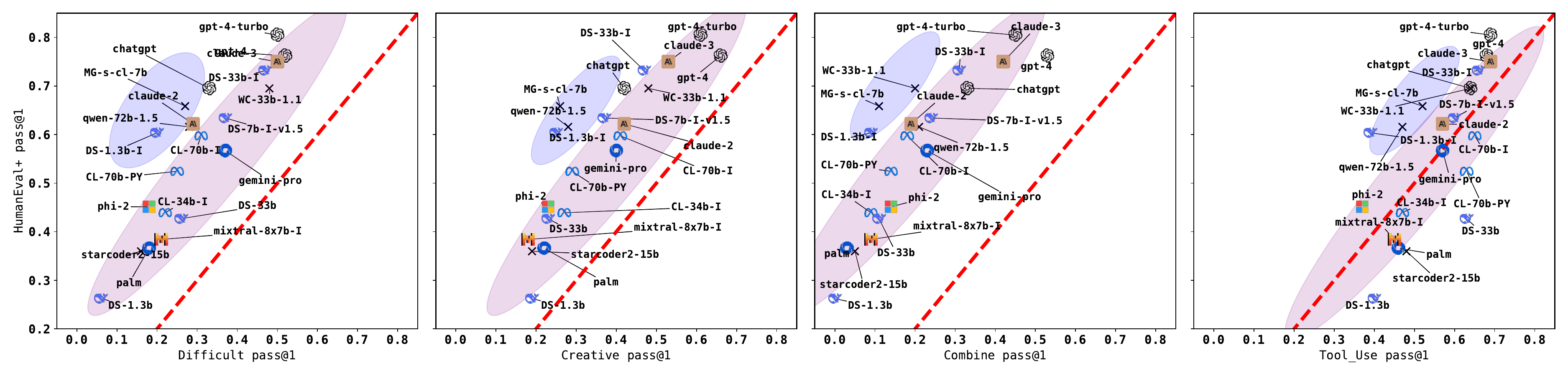}
    \centering
    \caption{Comparison of \passat{1} on \humaneval{+} and \tech datasets of selected models. 
    The \textcolor{red}{red} dotted identity line (\ie $x = y$) represents equivalent performance on both \humaneval and \tech. 
    For each benchmark, we cluster the \llm{s} into 1) \textcolor{linkcolor}{purple} region -- aligned performance on \humaneval and \tech and 2) \textcolor{blue}{blue} region -- over performant \llm{s} on \humaneval compared with \tech results.}
    \label{fig:overfit}
\end{figure}

\begin{figure*}
    \centering
    \begin{subfigure}[b]{0.48\textwidth}
    \centering
    \includegraphics[width=\textwidth]{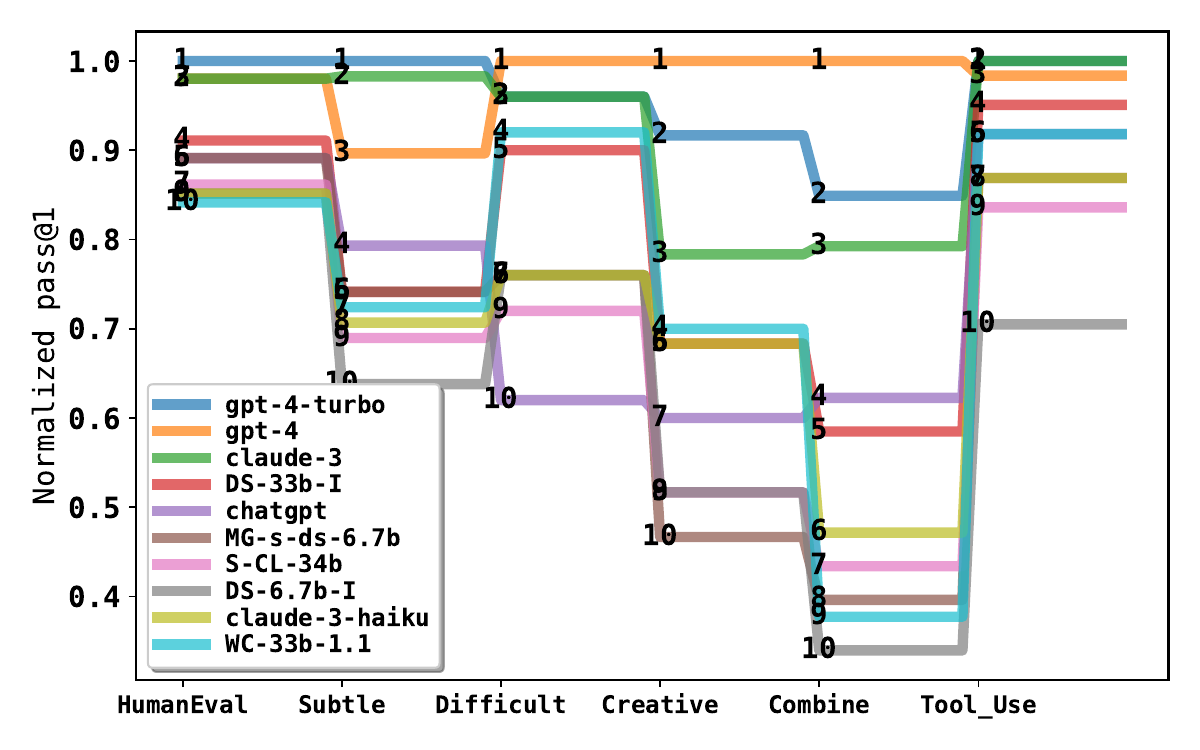}
    \caption{}
\end{subfigure}
\begin{subfigure}[b]{0.48\textwidth}
    \centering
    \includegraphics[width=\textwidth]{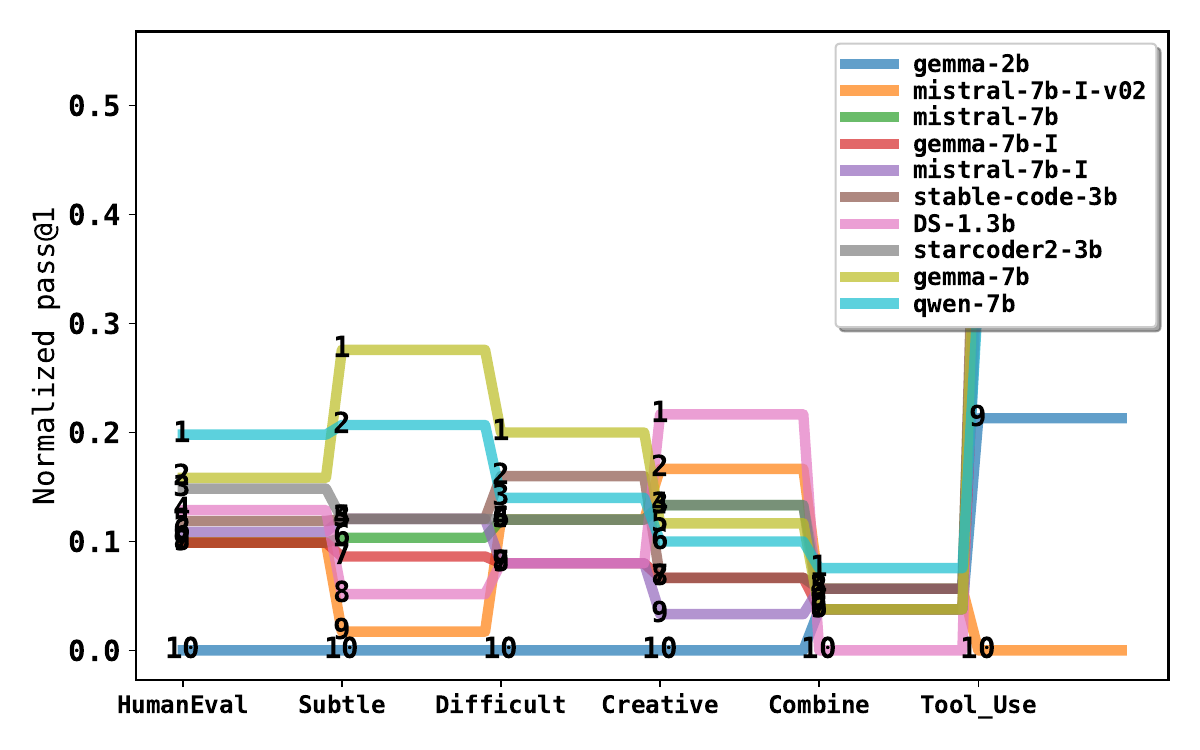}
    \caption{}
\end{subfigure}
\caption{Ranking changes across \tech benchmarks of 
(a) top-10 and (b) bottom-10 best performing \llm{s} on \humaneval respectively.
Y-axis shows the normalized \passat{1} score defined as the \llm \passat{1} normalized by the minimal and maximum \passat{1} achieved by all \llm{s} on benchmark.}
\label{fig:flow}
\end{figure*}

\begin{figure*}
    \centering
    \begin{subfigure}[b]{0.3\textwidth}
    \centering
    \includegraphics[width=\textwidth]{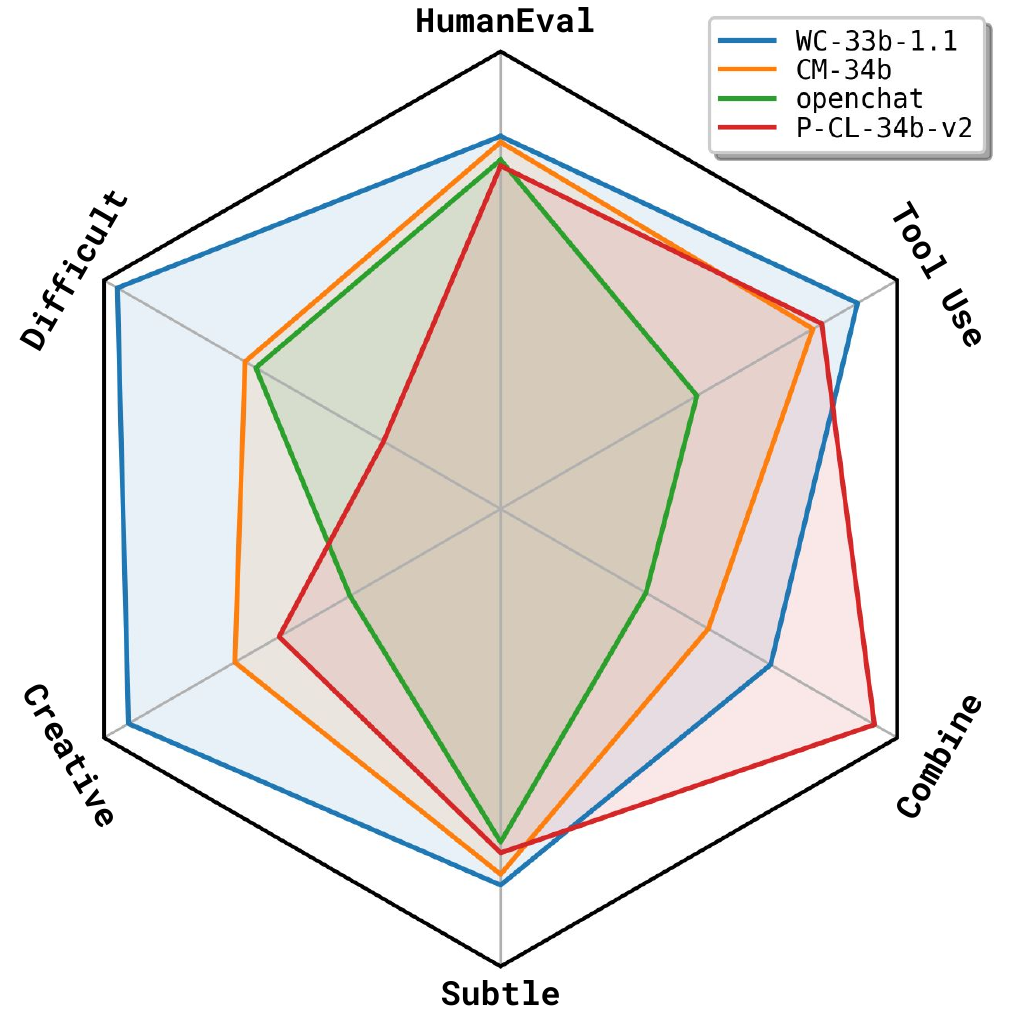}
    \caption{}
    \label{fig:radar_a}
\end{subfigure}
\begin{subfigure}[b]{0.3\textwidth}
    \centering
    \includegraphics[width=\textwidth]{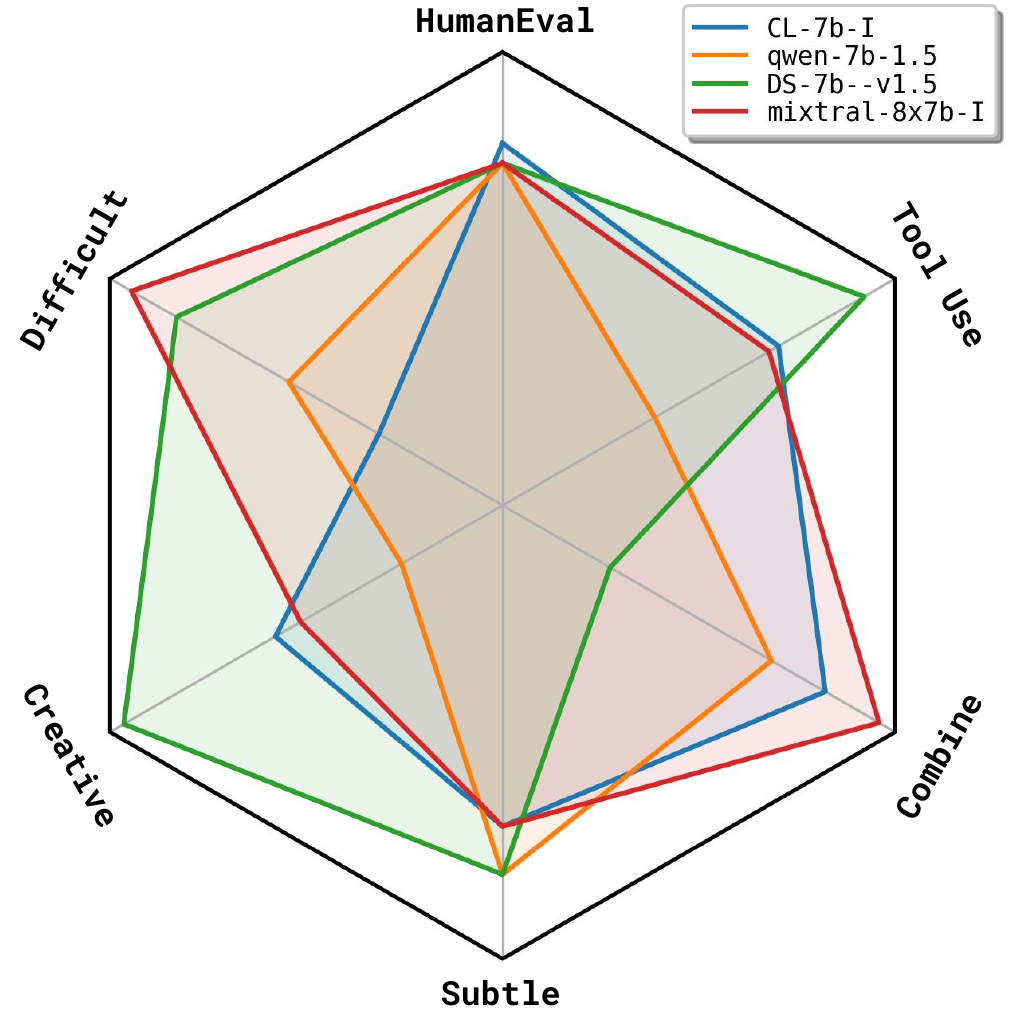}
    \caption{}
    \label{fig:radar_b}
\end{subfigure}
\caption{Radar graph of selected models with similar \humaneval scores.}
\label{fig:radar}
\end{figure*}

\parabf{\tech produces more complex and challenging benchmarks for program synthesis.}
Table~\ref{tab:main} shows the \passat{1} performance along with the ranking of \llm{s} on each of the semantic-altering \tech benchmarks with the average \passat{1} and ranking on all benchmarks in the last columns. 
First, compared to the success rate on \humaneval, when evaluated on \tech, all \llm{s} \textbf{consistently perform worse}.
For example, the state-of-the-art \gptfour, \gptfour-{Turbo} and \claude{-3} models solve close to 85\% of all \humaneval problems but fall almost below 50\% \passat{1} when evaluated on the \eedifficult problems. 
On average, across all benchmarks, the performance of \llm{s} decreased by \totaldecrease  (\eedifficult: \difficultdecrease, \eecreative: \creativedecrease, \eesubtle: \subtledecrease, \eecombine: \combinedecrease, and \eetoolusing: \toolusingdecrease). 
Additionally, this drop is not uniform across all \llm{s} and can range from \mindecrease to \maxdecrease.

\parabf{\llm{s} struggles on \tech benchmarks compare to high performance achieved on \humaneval.}
One surprising finding is that, on \eesubtle, where only small changes are made to original problem with the roughly the same level of difficulty, the average performance of \llm{s} drops by \subtlecomparisondecrease{} across the same 100 problems.
It is important to note that, as the \passat{1} score is generally higher on the first 100 problems than the complete 164 \humaneval problems, this back-to-back performance drop is much higher than the performance drop from \humaneval to \eesubtle mentioned above (which is \subtledecrease).
Furthermore, we can also identify \llm{s} which struggle heavily on specific types of problems compared to their relative performance on \humaneval. 
Figure~\ref{fig:overfit} shows scatter plot of \humaneval{+} and \tech scores of selected \llm{s}. 
As we saw before, the significant portions of the models tends to be worse on \tech than \humaneval (\ie purple shaded region).
However, there exists \llm{s} that have a \textit{much} higher \humaneval score compared to their performance on \tech (\ie blue shaded region).
This highlights potential data leakage of popular benchmarks where \llm performances are artificially inflated but do not translate to more difficult or other program synthesis problems.

\parabf{Significant ranking changes of \llm{s} across different \tech benchmarks.} 
In Figure~\ref{fig:flow},
compared to the existing parity -- where top models all perform similarly on \humaneval, we observe drastic differences in ranking changes on \tech.
We observe that while the relative difference between the top 5 models on \humaneval is less than 10\%, the difference on \tech on average is over 20\%.
Due to such saturation in top model performance, existing benchmarks may not reliably rank the program synthesis ability of each model. 
Taking a closer look at specific models, while \claude{-3} and \gptfour are tied for the 2nd best \humaneval score, they both excel at different types of problems: \gptfour performs best on difficult and creative problems while \claude{-3} can better reason about helper functions in \eetoolusing and are less affected by subtle changes from original \humaneval. 
Furthermore, while \gptfour{}-{Turbo} achieves the top \humaneval and \humaneval{+} score, it falls off compare to the base \gptfour variant where it is worse on \eedifficult, \eecreative and \eecombine problems. 
Such evaluation cannot be gained through naively reporting existing coding benchmark performance.
Overall, by evolving the original benchmark into more difficult and diverse problems of different types, \tech can provide a more holistic evaluation and ranking of the coding ability of \llm{s}.

\parabf{\tech can be used to comprehensively compare multiple models.}
Figure~\ref{fig:radar} shows two radar graphs of two sets of \llm{s}.
In Figure~\ref{fig:radar_a}, while both \wizardcoder-{1.1} and \phindllamatwo are top performing \llm{s} and have similar \humaneval scores, they perform drastically differently across the benchmarks in \tech.
\wizardcoder-{1.1} is better on \eedifficult and \eecreative and \phindllamatwo are better on \eecombine problems. 
This can be partially explained through the training dataset used in each \llm where \wizardcoder-{1.1} uses an evolving dataset to generate more complex and difficult problems whereas \phindllamatwo is fine-tuned on high quality programming problems that seems to boost the ability to solve programs which combines multiple smaller programming concepts. 
Similar phenomenon can also be observed in Figure~\ref{fig:radar_b}. 
Different from just reporting a singular \passat{k} score, \tech also allows detailed analysis across the different dimension of coding capability to identify particular domains or type of synthesis questions the \llm struggles or excels in.

\begin{figure}
    \includegraphics[width=\columnwidth]{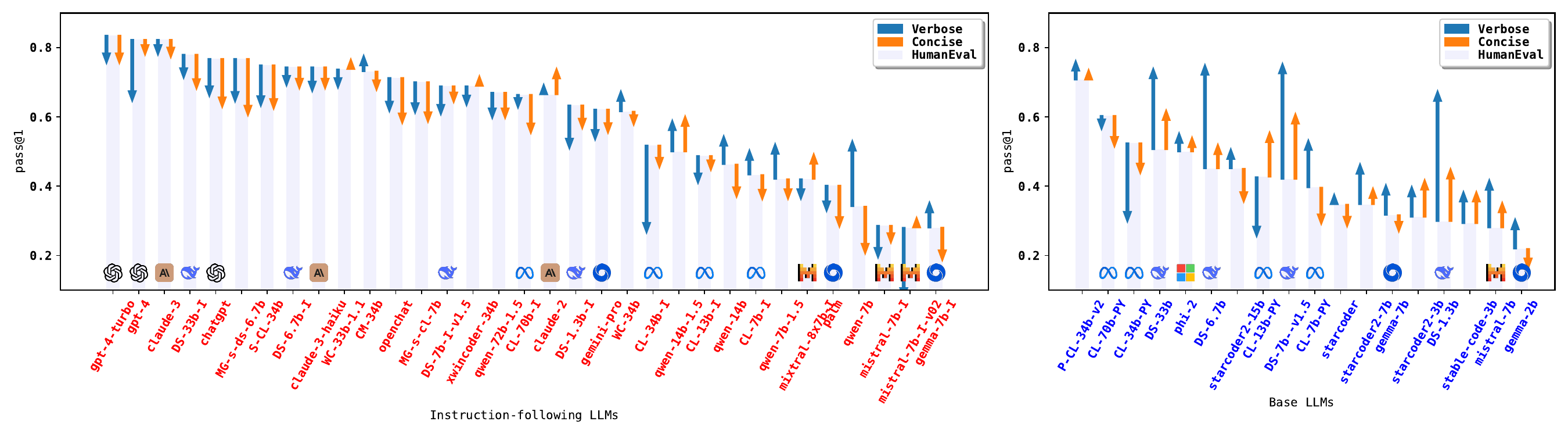}
    \centering
    \caption{\humaneval \passat{1} separated into instruction-following and non-instruction-following \llm{s} with relative decrease or increase in \passat{1} on \eeverbose and \eecombine.}
    \label{fig:semantic_equivalent_drop}
\end{figure}

\parabf{Instruction-following \llm{s} are sensitive to subtle or rephrasing of problem docstring.}
Unlike the semantic-altering benchmarks in \tech, 
the semantic-preserving problems do not always lead to a decrease in performance. 
Figure~\ref{fig:semantic_equivalent_drop} shows the \humaneval score (bar) and the relative performance drop or improvement (arrows) on \eeverbose and \eeconcise separated into instruction-following and base \llm{s}.
We observe that almost all instruction-following \llm{s} drops in performance (on average 3.4\% and 4.0\% decrease on \eeverbose and \eeconcise respectively) when evaluated on the two semantic-preserving dataset compared to the original \humaneval. 
This is drastically different from the non-instruction-following variants where we even observe performance improvements (on average 0.5\% and 2.1\% increase on \eeverbose and \eeconcise respectively).
\eeverbose and \eeconcise do not change the semantic meaning of the original problem except reword it in either a more verbose or concise manner. 
Prior work~\cite{deng2023rephrase} has shown that by smartly rephrasing the original problem description, one can further boost \llm performance and we observe the similar phenomenon here mostly only for non-instruction-following models.
This further points to possibility of overfitting to the exact descriptions utilized in \humaneval especially for instruction-tuned \llm{s}.

Additionally, even on the semantic-altering benchmark of \eesubtle, where only subtle changes to the original problem are applied, on average, instruction-following \llm{s} drops by 7.6\% whereas base models only decreases by less than 1\% relative to their \humaneval performance. 
These findings across \llm types show that while 
instruction-tuning is expected to align better with detailed task instructions, it fails to distinguish between these subtle changes in docstring, indicating potential memorization or contamination of prior evaluation benchmarks.

\subsection{Problem Composition}
\label{sec:compose_general}

\begin{table}
\setlength{\tabcolsep}{2pt}
  \caption{Detailed results of top performing \llm{s} on \eecombine and \eecombine-\textsc{Naive}. 
  ``\humaneval'' is categorized into ``\textit{pass both}'', ``\textit{pass one}'' and ``\textit{pass none}'', depending on the success on the two parent problems used to create \eecombine and \eecombine-\textsc{Naive}.
  ``\eecombine (Solved)'' and ``\eecombine-\textsc{Naive} (Solved)'' then show the distribution of successfully solved problems in the composition dataset that came from the previous categories.
``\textit{Composition Percentage}'' is defined as the percentage of ``\textit{pass both}'' problems the \llm can \emph{still} solve when combining both of these problems. 
  }\label{tab:combine}
  \centering
\scriptsize
\begin{tabular}{l c ccc|cccc}
\toprule
& \multirow{2}*{Size} & \multicolumn{3}{c}{\humaneval} & \multicolumn{3}{c}{\eecombine (Solved)} & \multirow{2}*{\makecell{Composition\\Percentage}}\\ 
\cmidrule(lr){3-5}
\cmidrule(lr){6-8}
& & pass both & pass one & pass none & pass both & pass one & pass none & \\

\midrule
\openai{ }\gptfour{\speech} & NA & \cellcolor[HTML]{f3f3ff}93 & \cellcolor[HTML]{f3f3ff}7 & \cellcolor[HTML]{f3f3ff}0 & \cellcolor[HTML]{ffe1ee}50 &\cellcolor[HTML]{ffe1ee}3 & \cellcolor[HTML]{ffe1ee}0& \textbf{\textcolor{red}{53.8\%}} \\
\openai{ }\gptfour{-Turbo}{\speech} & NA & \cellcolor[HTML]{f3f3ff}79 & \cellcolor[HTML]{f3f3ff}19 & \cellcolor[HTML]{f3f3ff}2 & \cellcolor[HTML]{ffe1ee}38 &\cellcolor[HTML]{ffe1ee}6 & \cellcolor[HTML]{ffe1ee}1& \textbf{\textcolor{red}{48.1\%}} \\
\anthropic{ }\claude{-3}{\speech} & NA & \cellcolor[HTML]{f3f3ff}81 & \cellcolor[HTML]{f3f3ff}19 & \cellcolor[HTML]{f3f3ff}0 & \cellcolor[HTML]{ffe1ee}35 &\cellcolor[HTML]{ffe1ee}7 & \cellcolor[HTML]{ffe1ee}0& \textbf{\textcolor{red}{43.2\%}} \\
\openai{ }\chatgpt{\speech} & NA & \cellcolor[HTML]{f3f3ff}65 & \cellcolor[HTML]{f3f3ff}34 & \cellcolor[HTML]{f3f3ff}1 & \cellcolor[HTML]{ffe1ee}24 &\cellcolor[HTML]{ffe1ee}9 & \cellcolor[HTML]{ffe1ee}0& \textbf{\textcolor{red}{36.9\%}} \\
\deepseeklogo{ }\deepseekinstruct{\speech} & 33b & \cellcolor[HTML]{f3f3ff}71 & \cellcolor[HTML]{f3f3ff}27 & \cellcolor[HTML]{f3f3ff}2 & \cellcolor[HTML]{ffe1ee}29 &\cellcolor[HTML]{ffe1ee}2 & \cellcolor[HTML]{ffe1ee}0& \textbf{\textcolor{red}{40.8\%}} \\
\anthropic{ }\claude{-3}{-haiku}{\speech} & NA & \cellcolor[HTML]{f3f3ff}63 & \cellcolor[HTML]{f3f3ff}34 & \cellcolor[HTML]{f3f3ff}3 & \cellcolor[HTML]{ffe1ee}19 &\cellcolor[HTML]{ffe1ee}6 & \cellcolor[HTML]{ffe1ee}0& \textbf{\textcolor{red}{30.2\%}} \\
\deepseeklogo{ }\deepseekvonefive{-Inst}{\speech} & 7b & \cellcolor[HTML]{f3f3ff}62 & \cellcolor[HTML]{f3f3ff}37 & \cellcolor[HTML]{f3f3ff}1 & \cellcolor[HTML]{ffe1ee}18 &\cellcolor[HTML]{ffe1ee}6 & \cellcolor[HTML]{ffe1ee}0& \textbf{\textcolor{red}{29.0\%}} \\
\google{ }\gemini{\speech} & NA & \cellcolor[HTML]{f3f3ff}46 & \cellcolor[HTML]{f3f3ff}45 & \cellcolor[HTML]{f3f3ff}9 & \cellcolor[HTML]{ffe1ee}19 &\cellcolor[HTML]{ffe1ee}2 & \cellcolor[HTML]{ffe1ee}2& \textbf{\textcolor{red}{41.3\%}} \\
\midrule

& & \multicolumn{3}{c}{\humaneval} & \multicolumn{3}{c}{\eecombine-\textsc{naive} (Solved)} & \\ 
\cmidrule(lr){3-5}
\cmidrule(lr){6-8}

\openai{ }\gptfour{\speech} & NA & \cellcolor[HTML]{f3f3ff}1018 & \cellcolor[HTML]{f3f3ff}55 & \cellcolor[HTML]{f3f3ff}1 & \cellcolor[HTML]{ffe1ee}766 &\cellcolor[HTML]{ffe1ee}7 & \cellcolor[HTML]{ffe1ee}0& \textbf{\textcolor{red}{75.2\%}} \\
\openai{ }\gptfour{-Turbo}{\speech} & NA & \cellcolor[HTML]{f3f3ff}863 & \cellcolor[HTML]{f3f3ff}195 & \cellcolor[HTML]{f3f3ff}16 & \cellcolor[HTML]{ffe1ee}407 &\cellcolor[HTML]{ffe1ee}61 & \cellcolor[HTML]{ffe1ee}3& \textbf{\textcolor{red}{47.2\%}} \\
\anthropic{ }\claude{-3}{\speech} & NA & \cellcolor[HTML]{f3f3ff}796 & \cellcolor[HTML]{f3f3ff}268 & \cellcolor[HTML]{f3f3ff}10 & \cellcolor[HTML]{ffe1ee}359 &\cellcolor[HTML]{ffe1ee}96 & \cellcolor[HTML]{ffe1ee}1& \textbf{\textcolor{red}{45.1\%}} \\
\openai{ }\chatgpt{\speech} & NA & \cellcolor[HTML]{f3f3ff}799 & \cellcolor[HTML]{f3f3ff}261 & \cellcolor[HTML]{f3f3ff}14 & \cellcolor[HTML]{ffe1ee}474 &\cellcolor[HTML]{ffe1ee}79 & \cellcolor[HTML]{ffe1ee}1& \textbf{\textcolor{red}{59.3\%}} \\
\deepseeklogo{ }\deepseekinstruct{\speech} & 33b & \cellcolor[HTML]{f3f3ff}740 & \cellcolor[HTML]{f3f3ff}304 & \cellcolor[HTML]{f3f3ff}30 & \cellcolor[HTML]{ffe1ee}462 &\cellcolor[HTML]{ffe1ee}95 & \cellcolor[HTML]{ffe1ee}5& \textbf{\textcolor{red}{62.4\%}} \\
\anthropic{ }\claude{-3}{-haiku}{\speech} & NA & \cellcolor[HTML]{f3f3ff}592 & \cellcolor[HTML]{f3f3ff}409 & \cellcolor[HTML]{f3f3ff}73 & \cellcolor[HTML]{ffe1ee}286 &\cellcolor[HTML]{ffe1ee}133 & \cellcolor[HTML]{ffe1ee}17& \textbf{\textcolor{red}{48.3\%}} \\
\deepseeklogo{ }\deepseekvonefive{-Inst}{\speech} & 7b & \cellcolor[HTML]{f3f3ff}634 & \cellcolor[HTML]{f3f3ff}372 & \cellcolor[HTML]{f3f3ff}68 & \cellcolor[HTML]{ffe1ee}393 &\cellcolor[HTML]{ffe1ee}130 & \cellcolor[HTML]{ffe1ee}17& \textbf{\textcolor{red}{62.0\%}} \\
\google{ }\gemini{\speech} & NA & \cellcolor[HTML]{f3f3ff}364 & \cellcolor[HTML]{f3f3ff}535 & \cellcolor[HTML]{f3f3ff}175 & \cellcolor[HTML]{ffe1ee}225 &\cellcolor[HTML]{ffe1ee}205 & \cellcolor[HTML]{ffe1ee}37& \textbf{\textcolor{red}{61.8\%}} \\

\bottomrule

\end{tabular}
\end{table}

\parabf{Composition problems.} 
The ability to compose different known concepts to solve new problems is known as \textit{compositional generalization}~\cite{keysers2020measuring}.
This skill is essential for code synthesis, especially for complex problems in real-world programs. 
However, measuring compositional generalization in \llm{} presents a fundamental challenge since it requires controlling the relationship between training and test distributions~\cite{shi2024exedec}.
While it is not easy to control the pre-training data of \llm{s}, we have more control in the testing phase.
Hence, we focus on program concepts that have been demonstrated to fall within the capabilities of an \llm{}, and explore whether this proficiency extends to the combination of program concepts.
As such, we start by taking a deeper look at the \eecombine problems evolved from combining previous \humaneval problems.

First half of Table~\ref{tab:combine} shows the detailed breakdown of the \eecombine dataset results on the top 8 performing \llm{s}. 
We observe that almost all problems solved in \eecombine came from the pass both category, which is intuitive as we do not expect \llm{s} to solve a problem composed of subproblems that it cannot already solve.
However, we see that overall, the composition percentage is quite low as only \gptfour is able achieve greater than half.%
This demonstrates, for the first time, that while state-of-the-art \llm{s} can achieve a high pass rate on simple programming tasks in general-purpose languages like Python, they still struggle with generalizing and composing these known concepts to address more complex problems. 

\begin{wrapfigure}{r}{0.4\textwidth}
  \begin{center}
    \includegraphics[width=0.4\textwidth]{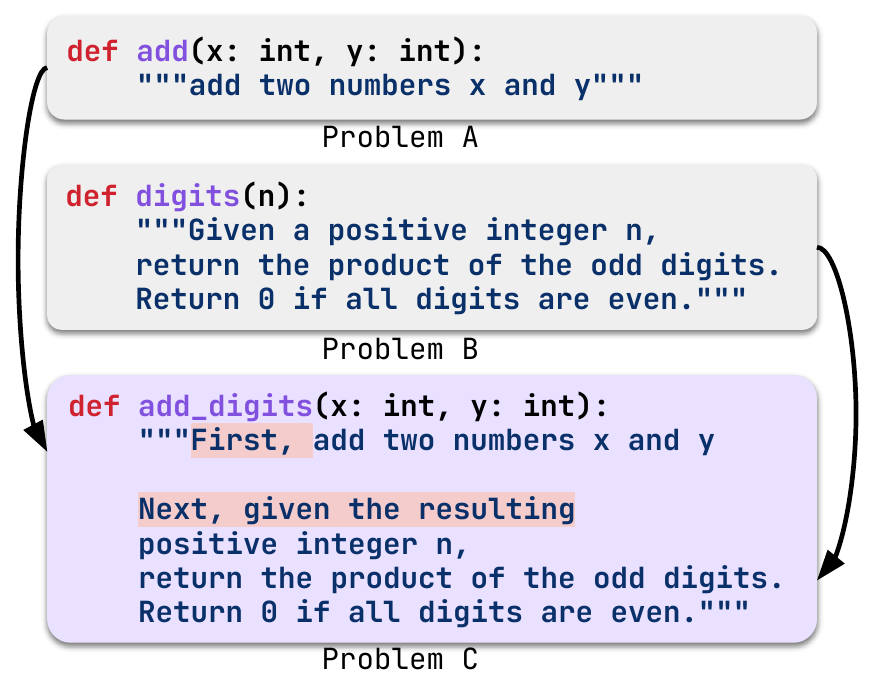}
  \end{center}
  \caption{\eecombine-\textsc{naive} problem}
  \label{fig:naive_combine_example}
\end{wrapfigure}

\parabf{Naive combination problems.} Since \eecombine problems are not guaranteed to not contain additional new logic or concepts, we build a simplified dataset for sequential composition. 
Let $A$ and $B$ be two separate problems with $x$ as input(s) for $A$, we aim to create a new problem $C$ with same inputs where the solution can be written as $B(A(x))$.
To accomplish this, the new problem includes a sequential docstring by attaching the docstring of problem $A$ followed by $B$. 
Directly concatenating them will lead to unclear descriptions, as such, for each problem in \humaneval, we manually create two separate variants based on which order the problem may come in the new docstring. 
Figure~\ref{fig:naive_combine_example} shows an example naive combination problem with the manual sequential instruction highlighted in red. 
Using these modified problem docstrings, we build a sequential combination dataset -- \eecombine-\textsc{naive}, containing 1074 problems by randomly combining problems filtering for input output matching (\ie type of $A(x)$ should equal to type of $y$ in $B(y)$)

The latter half of Table~\ref{tab:combine} shows the results on \eecombine-\textsc{naive} following the same setup as \eecombine. 
We observe that while the composition percentage on the naive dataset improves significantly compared to the evolved \eecombine dataset, it still fails to reach near perfection, with the best \llm being able to only solve 3/4 of prior pass both problems.
While existing training or inference paradigms for \llm{s} for code focus on obtaining high quality datasets boosted with instruction-tuning,
our result shows that existing \llm{s} still struggle with the concept of problem composition to tackle more complex problems.
We hope future research can design novel training methods to tackle this limitation. 

\subsection{Problem Decomposition}
\label{sec:decompose}

\newcommand\nocell[1]{\multicolumn{#1}{c|}{}}

\begin{table}
\setlength{\tabcolsep}{2pt}
  \caption{Detailed results of top performing \llm{s} on \eedecompose.
  ``\humaneval'' shows the pass/fail breakdown of the 50 seed \humaneval problems. 
Each of these 50 problems, initially pass or failed, is decomposed into two subproblems. These are further categorized into ``\textit{pass both}'', ``\textit{pass one}'' and ``\textit{pass none}'', based on whether the \llm can solve both subproblems.
``\textit{Decomp. \%}'' is the percentage of originally passing problems for which the \llm can solve both decomposed subproblems.
Similarly, ``\textit{Recomp. \%}'' is the percentage of originally failing problems for which the \llm can solve both decomposed subproblems.
}\label{tab:decompose}
  \centering
\scriptsize
\begin{tabular}{l c| cc|ccc|ccc|ccc}
\toprule
& \multirow{2}*{Size} & \multicolumn{2}{c}{\multirow{2}{*}{\humaneval}} & \multicolumn{6}{|c|}{\eedecompose} & \multirow{3}*{\makecell{Decomp.\\ \%}} & \multirow{3}*{\makecell{Recomp.\\ \%}}\\ 
\cmidrule(lr){5-10}
& & & & \multicolumn{3}{c|}{\humaneval pass} & \multicolumn{3}{c|}{\humaneval fail} \\
\cmidrule(lr){3-4}
\cmidrule(lr){5-8}
\cmidrule(lr){8-10}
& & pass & fail & both pass & one pass & both fail & both pass & one pass & both fail & & \\
\midrule
\openai{ }\gptfour{\speech} & NA & \cellcolor[HTML]{f3f3ff}47 & \cellcolor[HTML]{f3f3ff}3 &37 &10 & 0& 0 &3 & 0& \textbf{\textcolor{red}{78.7\%}} & \textbf{\textcolor{blue}{0.0\%}} \\
\openai{ }\gptfour{-Turbo}{\speech} & NA & \cellcolor[HTML]{f3f3ff}39 & \cellcolor[HTML]{f3f3ff}11 &29 &9 & 1& 4 &6 & 1& \textbf{\textcolor{red}{74.4\%}} & \textbf{\textcolor{blue}{36.4\%}} \\
\anthropic{ }\claude{-3}{\speech} & NA & \cellcolor[HTML]{f3f3ff}39 & \cellcolor[HTML]{f3f3ff}11 &26 &11 & 2& 6 &5 & 0& \textbf{\textcolor{red}{66.7\%}} & \textbf{\textcolor{blue}{54.5\%}} \\
\openai{ }\chatgpt{\speech} & NA & \cellcolor[HTML]{f3f3ff}33 & \cellcolor[HTML]{f3f3ff}17 &19 &13 & 1& 11 &4 & 2& \textbf{\textcolor{red}{57.6\%}} & \textbf{\textcolor{blue}{64.7\%}} \\
\deepseeklogo{ }\deepseekinstruct{\speech} & 33b & \cellcolor[HTML]{f3f3ff}33 & \cellcolor[HTML]{f3f3ff}17 &18 &14 & 1& 8 &9 & 0& \textbf{\textcolor{red}{54.5\%}} & \textbf{\textcolor{blue}{47.1\%}} \\
\anthropic{ }\claude{-3}{-haiku}{\speech} & NA & \cellcolor[HTML]{f3f3ff}28 & \cellcolor[HTML]{f3f3ff}22 &16 &10 & 2& 11 &11 & 0& \textbf{\textcolor{red}{57.1\%}} & \textbf{\textcolor{blue}{50.0\%}} \\
\deepseeklogo{ }\deepseekvonefive{-Inst}{\speech} & 7b & \cellcolor[HTML]{f3f3ff}27 & \cellcolor[HTML]{f3f3ff}23 &18 &8 & 1& 9 &11 & 3& \textbf{\textcolor{red}{66.7\%}} & \textbf{\textcolor{blue}{39.1\%}} \\
\google{ }\gemini{\speech} & NA & \cellcolor[HTML]{f3f3ff}19 & \cellcolor[HTML]{f3f3ff}31 &13 &6 & 0& 10 &18 & 3& \textbf{\textcolor{red}{68.4\%}} & \textbf{\textcolor{blue}{32.3\%}} \\

\bottomrule

\end{tabular}
\end{table}

Given our analysis and benchmark on combining different problems together, a nature follow-up would be to look at \textit{problem decomposition} -- decomposing larger problems into multiple subproblems.
We start by selecting 50 \humaneval problems and then follow our approach in Section~\ref{sec:approach} to decompose each original problem into two smaller subproblems, creating 100 problems in our \eedecompose benchmark. 

Table~\ref{tab:decompose} shows the results of selected \llm{s} on \eedecompose (the same set of \llm{s} as \eecombine). 
We first observe that similar to the composition percentage in the \eecombine and \eecombine-\textsc{naive} problems, \llm{s} do not achieve a high decomposition percentage.
One possible interpretation is that current \llm{s} are trained to memorize or recover seen outputs in their training data, and when used for program synthesis, they cannot generalize the concepts from training data.
This is demonstrated by not being able to solve smaller subproblems obtained from solved more difficult parent problems. 
On the other hand, we show that \llm{s} can sometimes solve both smaller subproblems even when the original parent problem is not solved (\ie recomposition percentage).
\eedecompose is akin to breaking the harder problem down into easier subproblems, which is related to planning in prior work~\cite{jiang2023self}.
We hope future work can again build on these insights to achieve the best of both worlds in being able to succcesfully generalize difficult concepts into subproblems and adopting decomposing/planning to solve additional challenging problems.

\subsection{Tool Using}
\label{sec:tool_using}

\begin{figure}
    \includegraphics[width=\columnwidth]{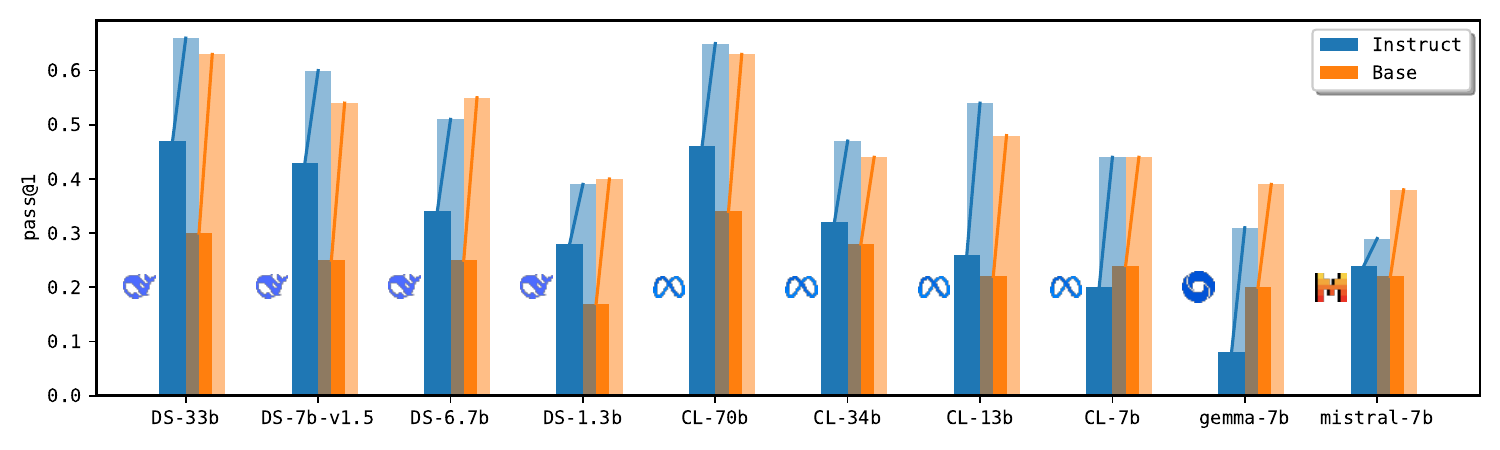}
    \centering
    \caption{\passat{1} improvement from \eetoolusing-\textsc{Main\_Only} to \eetoolusing on selected instruction-following models and their base variants.}
    \label{fig:tool_using_compare}
\end{figure}

\newcommand{\toolusinginstructincrease}{60.4\%\xspace}
\newcommand{\toolusingnoneincrease}{122.0\%\xspace}
\newcommand{\toolusingincrease}{81.3\%\xspace}
\newcommand{\toolusingmainonlypass}{28.6\%\xspace}

We further analyze the \eetoolusing dataset, which contains pre-defined helper or auxiliary functions in addition to the main synthesis problem. 
Additionally, we construct \eetoolusing-\textsc{Main\_Only} dataset, which contains the same set of problem as \eetoolusing, except that the input to the \llm consists only of the main problem description without including any helpers. 
Using both datasets together, we can evaluate the ability of \llm{s} to use helper functions to solve more complex problem. 
We observe that compared to scenarios without any helper functions (average \passat{1} of \toolusingmainonlypass), \llm{s} on average improve by \toolusingincrease when provided with the helper functions.
This is to be expected as the helper functions provides additional utilities in aiding to solve the more complex problem.
However, this improvement is not uniform, as we see that the average improvement when given the auxiliary functions for instruction-following models is only \toolusinginstructincrease compared to the non-instruction-following \llm{s}' improvement of \toolusingnoneincrease. 

Figure~\ref{fig:tool_using_compare} show the detailed comparison between 10 instruction-following and their base \llm{s} on both the \eetoolusing-\textsc{Main\_Only} and \eetoolusing dataset. 
We observe that without the helpers, the instruction-following models significantly outperform their base \llm{s}.
However, once the helpers are provided, this gap is drastically decreased, with cases even where the base models outperform their instruction-following counterparts. 
As real-world coding involves understanding, using, and then reusing existing functions across different places in the repository, being able to successfully leverage auxiliary methods is key.
Current instruction-following \llm{s} are generally fine-tuned with data consisting of self-contained code snippets without the interaction and learning of function usages.
This is further exacerbated by prior benchmarks, which mostly use self-contained functions, thus cannot expose the insufficient tool-using capability of such models. 
In \tech, with \eetoolusing and \eetoolusing-\textsc{Main\_Only}, we demonstrate this gap in evaluation and hope to inspire future research on this important aspect of code \llm{s}.

\section{Related Work}

\parabf{Large language models for code.} 
Starting with the general development of \llm{s} for general purpose tasks, developers have applied \llm{s} to perform code-related tasks by further training \llm{s} using collected code snippets from open-source repositories.
Such \llm{s} include \codex~\cite{codex}, PolyCoder~\cite{polycoder}, CodeT5~\cite{wang2021codet5}, \codegen~\cite{codegen}, InCoder~\cite{incoder}, \codellama~\cite{codellama}, StarCoder~\cite{starcoder}, StarCoder2~\cite{starcodertwo}, \deepseek~\cite{deepseek}, etc. 
These \llm{s} can autoregressive complete code given the relevant prefix (\eg docstrings for function completion).
More recently, following the advancement in NLP, researchers have applied instruction-tuning methods to train code-specific \llm{s} that are well-versed in following instructions.
Examples of such \llm{s} include \codellamainstruct~\cite{codellama} and \deepseekinstruct~\cite{deepseek}. 
\wizardcoder~\cite{wizardcoder} instruction-tunes the model using Evol-Instruct to create more complex instructions.
\magicoder~\cite{magicoder} develops OSS-Instruct by synthesizing high quality instruction data from open-source code snippets.
OpenCodeInterpreter~\cite{zheng2024opencodeinterpreter} additionally leverages execution feedback for instruction-tuning in order to better support multi-turn code generation and refinement. 

\parabf{Program synthesis benchmarking.} 
\humaneval~\cite{codex} and \mbpp~\cite{austin2021program} are two of the most widely-used handcrafted code generation benchmarks complete with test cases to check for the correctness of \llm outputs.
Building on these popular benchmarks, additional variants have been crafted including: \humaneval{+}~\cite{evalplus} which improves the two benchmarks with more complete testcases; 
\humaneval-{X}~\cite{zheng2023codegeex} which extends \humaneval to C++, Javascript and Go;
MultiPL-E~\cite{cassano2023multipl} which further extends both \humaneval and \mbpp to 18 coding languages.
Similarly, other benchmarks have been developed for specific domains: DS-1000~\cite{ds1000} and Arcade~\cite{arcade} for data science APIs; 
ODEX~\cite{wang2023execution} for open-domain code generation covering a diverse range of libraries;
CodeContests~\cite{codecontest}, \apps~\cite{hendrycksapps2021} and LiveCodeBench~\cite{livecodebench} for programming contests; ClassEval~\cite{classeval} for class-level generations, and SWE-Bench~\cite{swebench} for real-world software engineering tasks.
Different from prior benchmarks which require handcraft problems from scratch -- high manual effort or scrape open-source repositories or coding contest websites -- leading to unavoidable data leakage, 
\tech directly uses \llm{s} to \textit{evolve} existing benchmark problems to create new complex evaluation problems.
Furthermore, contrasting with the narrow scope of prior benchmarks (often focusing on a single type or problem, i.e., coding contests), \tech utilizes targeted transformation to evolve problems into different domains, allowing for a more holistic evaluation of program synthesis using \llm{s}.

\section{Conclusion}

We present \tech -- a set of program synthesis benchmarks created by \textit{evolving} existing problems into different target domains.
We build on top of the popular \humaneval benchmark to produce 828 problems across 7 different benchmarks for a holistic and comprehensive evaluation of \llm program synthesis ability.
Our results on \textbf{\totalmodels} \llm{s} show that compare to high performance on standard benchmarks, there is drastic drop in performance (on average \totaldecrease) when evaluated on \tech.
Additionally, we observe significant ranking differences compared to previous leaderboards, indicating potential overfitting of popular \llm{s} on existing benchmarks.
Throughout the paper, we provide additional insights, including the brittleness of instruction-following \llm{s} as well as problem composition and decomposition abilities. 
We hope \tech not only provides a valuable benchmarking suite for program synthesis but also inspires future code \llm builders to recognize the shown limitations of existing code \llm{s} and develop novel and targeted training approaches for code.
We have open-sourced the \tech benchmarks, tools, and complete \llm generations available at \textcolor{linkcolor}{\url{https://github.com/evo-eval/evoeval}}

\section{Acknowledgment}

We thank Owen Colegrove for his help on starting this project and providing valuable feedback throughout, Jiawei Liu for providing helpful discussions and Yifeng Ding for his help in running experiments.

\bibliography{reference}
\bibliographystyle{colm2024_conference}

\newpage
\appendix

\section{Evaluation \llm{s}}

\subsection{Evaluated \llm{s}} 

\begin{table}
\setlength{\tabcolsep}{2pt}
  \caption{Detailed overview of evaluated models. Model ID indicates either the API endpoint name or huggingface model name used for the particular model. Available Weights indicate whether the model is evaluated by accessing a close-sourced API endpoint or ran locally with provided weights. Note: \protect\speech denotes instruction-following \llm{s} 
  }
  \label{tab:models}
  \centering
    \scriptsize
\begin{tabular}{lr lr}
\toprule 

& Size & Model ID & Available Weights \\ 
\midrule
\openai{ }\gptfour{-Turbo}{\speech}\cite{gptfour} & NA & gpt-4-0125-preview & \xmark \\
\midrule
\openai{ }\gptfour{\speech}\cite{gptfour} & NA & gpt-4-0613 & \xmark \\
\midrule
\openai{ }\chatgpt{\speech}\cite{chatgpt} & NA & gpt-3.5-turbo-0125 & \xmark \\
\midrule
\anthropic{ }\claude{-3}{\speech}\cite{claudethree} & NA & claude-3-opus-20240229 & \xmark \\
\midrule
\anthropic{ }\claude{-3}{-haiku}{\speech}\cite{claudethree} & NA & claude-3-haiku-20240307 & \xmark \\
\midrule
\anthropic{ }\claude{-2}{\speech}\cite{claudetwo} & NA & claude-2.1 & \xmark \\
\midrule
\google{ }\gemini{\speech}\cite{gemini} & NA & gemini-1.0-pro & \xmark \\
\midrule
\google{ }\palm{\speech}\cite{palm} & NA & text-bison-001 & \xmark \\
\midrule
\multirow{3}*{\deepseeklogo{ }\deepseekinstruct{\speech}\cite{deepseek}} & 33b & deepseek-ai/deepseek-coder-33b-instruct &  \cmark\\
 & 6.7b & deepseek-ai/deepseek-coder-6.7b-instruct &  \cmark\\
 & 1.3b & deepseek-ai/deepseek-coder-1.3b-instruct &  \cmark\\
\midrule
\multirow{3}*{\deepseeklogo{ }\deepseek\cite{deepseek}} & 33b & deepseek-ai/deepseek-coder-33b-base &  \cmark\\
 & 6.7b & deepseek-ai/deepseek-coder-6.7b-base &  \cmark\\
 & 1.3b & deepseek-ai/deepseek-coder-1.3b-base &  \cmark\\
\midrule
\deepseeklogo{ }\deepseekvonefive{-Inst.}{\speech}\cite{deepseek} & 7b & deepseek-ai/deepseek-coder-7b-instruct-v1.5 &  \cmark\\
\midrule
\deepseeklogo{ }\deepseekvonefive\cite{deepseek} & 7b & deepseek-ai/deepseek-coder-7b-base-v1.5 &  \cmark\\
\midrule
\multirow{4}*{\meta{ }\codellamainstruct{\speech}\cite{codellama}} & 70b & codellama/CodeLlama-70b-Instruct-hf &  \cmark\\
 & 34b & codellama/CodeLlama-34b-Instruct-hf &  \cmark\\
 & 13b & codellama/CodeLlama-13b-Instruct-hf &  \cmark\\
 & 7b & codellama/CodeLlama-7b-Instruct-hf &  \cmark\\
\midrule
\multirow{4}*{\meta{ }\codellama\cite{codellama}} & 70b & codellama/CodeLlama-70b-Python-hf &  \cmark\\
 & 34b & codellama/CodeLlama-34b-Python-hf &  \cmark\\
 & 13b & codellama/CodeLlama-13b-Python-hf &  \cmark\\
 & 7b & codellama/CodeLlama-7b-Python-hf &  \cmark\\
\midrule
\wizardcoder{\speech}\cite{wizardcoder} & 34b & WizardLM/WizardCoder-Python-34B-V1.0 &  \cmark\\
\midrule
\wizardcoder{-1.1}{\speech}\cite{wizardcoder} & 33b & WizardLM/WizardCoder-33B-V1.1 &  \cmark\\
\midrule
\xwincoder{\speech}\cite{xwincoder} & 34b & Xwin-LM/XwinCoder-34B &  \cmark\\
\midrule
\phindllamatwo\cite{phindllamatwo} & 34b & Phind/Phind-CodeLlama-34B-v2 &  \cmark\\
\midrule
\codemillenials{\speech}\cite{codemillenials} & 34b & budecosystem/code-millenials-34b &  \cmark\\
\midrule
\speechlesscodellama{\speech}\cite{speechlesscodellama} & 34b & uukuguy/speechless-codellama-34b-v2.0 &  \cmark\\
\midrule
\magicoder{-s-DS}{\speech}\cite{magicoder} & 6.7b & ise-uiuc/Magicoder-S-DS-6.7B &  \cmark\\
\midrule
\magicoder{-s-CL}{\speech}\cite{magicoder} & 7b & ise-uiuc/Magicoder-S-CL-7B &  \cmark\\
\midrule
\multirow{3}*{\starcodertwo\cite{starcodertwo}} & 15b & bigcode/starcoder2-15b &  \cmark\\
 & 7b & bigcode/starcoder2-7b &  \cmark\\
 & 3b & bigcode/starcoder2-3b &  \cmark\\
\midrule
\starcoder\cite{starcoder} & 15b & bigcode/starcoder &  \cmark\\
\midrule
\mistrallogo{ }\mixtralinstruct{\speech}\cite{mixtral} & 8x7b & mistralai/Mixtral-8x7B-Instruct-v0.1 &  \cmark\\
\midrule
\mistrallogo{ }\mistralinstruct{-v02}{\speech}\cite{mistral} & 7b & mistralai/Mistral-7B-Instruct-v0.2 &  \cmark\\
\midrule
\mistrallogo{ }\mistralinstruct{\speech}\cite{mistral} & 7b & mistralai/Mistral-7B-Instruct-v0.1 &  \cmark\\
\midrule
\mistrallogo{ }\mistral\cite{mistral} & 7b & mistralai/Mistral-7B-v0.1 &  \cmark\\
\midrule
\openchat{\speech}\cite{openchat} & 7b & openchat/openchat-3.5-0106 &  \cmark\\
\midrule
\stablecode\cite{stablecode} & 3b & stabilityai/stable-code-3b &  \cmark\\
\midrule
\google{ }\gemma{-Inst.}{\speech\cite{gemma}} & 7b & google/gemma-7b-it &  \cmark\\
\midrule
\multirow{2}*{\google{ }\gemma\cite{gemma}} & 7b & google/gemma-7b &  \cmark\\
 & 2b & google/gemma-2b &  \cmark\\
\midrule
\microsoft{ }\phitwo\cite{phitwo} & 2.7b & microsoft/phi-2 &  \cmark\\
\midrule
\multirow{3}*{\qwen{\speech}\cite{qwenonefive}} & 72b & Qwen/Qwen1.5-72B-Chat &  \cmark\\
 & 14b & Qwen/Qwen1.5-14B-Chat &  \cmark\\
 & 7b & Qwen/Qwen1.5-7B-Chat &  \cmark\\
\midrule
\multirow{2}*{\qwenb{\speech}\cite{qwenone}} & 14b & Qwen/Qwen-14B-Chat &  \cmark\\
 & 7b & Qwen/Qwen-7B-Chat &  \cmark\\
 \bottomrule
\end{tabular}
\end{table}

Table~\ref{tab:models} shows the overview of the \totalmodels \llm{s} we evaluated in our work. 
For any \llm{s} which provide their open-source weights, we directly obtain them from huggingface model hub~\cite{HuggingFaceWebPage}\footnote{For certain \llm{s}, we may use the vLLM inference library for more efficient generation}.
For any close-sourced \llm{s}, we directly access their model endpoints using their providers.
For more detail on the access of each \llm, please check our repository: \textcolor{linkcolor}{\url{https://github.com/evo-eval/evoeval}}

\subsection{Detailed Evaluation Setup}

\begin{figure*}[h]
\centering
\begin{tcolorbox}
[enhanced,size=small,colback=black!5!white,flip title={interior hidden}]
\begin{lstlisting}[captionpos=b, breaklines=true, language=Python, escapechar=\%]
Please complete the following code snippet.
def transform_canvas(canvas: str) -> str:
    """
    You have an canvas containing either '#' (representing a wall), '-' (representing 
    an empty space), or 'P' (representing the point at which a painter starts). The painter 
    can move horizontally on the canvas and paints all empty spaces he encounters
    with '*' without crossing or hitting the walls.
    
    The task is to return an updated canvas with all the accessible spaces painted, 
    keeping wall configuration and unaccessible spaces same. If the canvas contains no painter 'P', 
    return the canvas as it is. If there are more than one 'P' or the number of painted space divides the empty spaces evenly, return 'Invalid canvas'.

    Examples:

    >>> transform_canvas('P----#-----#-----#-----')
    'P****#-----#-----#-----'
    
    >>> transform_canvas('--#-P#-----#-----#--#--')
    'Invalid canvas'
    
    >>> transform_canvas('-----#--P--#-----#-----')
    '-----#**P**#-----#-----'
    
    >>> transform_canvas('-----#-----#--P---#P----')
    'Invalid canvas'
    """
\end{lstlisting}
\end{tcolorbox}
\caption{Example input prompt for \gptfour{}}
\label{fig:app_example_gptinput}
\end{figure*}

\parabf{\llm generation.} As mentioned in Section~\ref{sec:method}, we report the \passat{1} score for each \llm on our dataset generated using greedy decoding (i.e., sampling with temperature = 0). 
For each \llm, we provide a specific input prompt depending on the model type. 
For base \llm{s} (i.e., not instruction-following variants), we use only the function headers as input. 
For instruction-following, we make the best effort to follow examples provided by each model maker on the exact instruction and format to use at the time of writing.
Specifically, for instruction-following \llm{s}, we ask the model to return the code snippet wrapped by code blocks (i.e., \`{}\`{}\`{}). 
Figure~\ref{fig:app_example_gptinput} shows an example input for \gptfour{} on a \eecreative problem. 

Furthermore, we also provide a custom sanitization script adopted from \evalplus~\cite{evalplus} which parses the raw \llm outputs for code block parsing (\eg removing \`{}\`{}\`{} indicators for instruction-following models) and end-of-string identifiers (\eg removing tokens like \CodeIn{</s>}). 
Each model generated output is passed into the sanitization script and the evaluation occurs on the sanitized outputs. 

\parabf{Oracle.} 
To evaluate the functional correctness of each \llm synthesized solution, we use differential testing by comparing the model output with the groundtruth output on a set of testcase inputs.
We build our evaluation framework on top of the \evalplus evaluation script used for \humaneval and \humaneval{+} benchmark which evaluates multiple problems and solutions in parallel for efficiency.  
For each testcase, we perform exact matching or check if the output is within an absolute difference threshold of $10^{-6}$ if the output is a floating point type. 
We additionally implement our evaluation script by recursively checking the type and performing the appropriate comparison (\eg dictionary outputs are first length checked for equivalence and then matching is done for each value and key). 
Furthermore, we also implement custom oracles for specific problems where there could be multiple solutions or simple tolerance or exact matching cannot fully guarantee correctness. 
We refer the reader again to our repository \textcolor{linkcolor}{\url{https://github.com/evo-eval/evoeval}} which contains the full implementation of each of our custom oracle. 
Additionally, we also use timeout as another evaluation method. Our setting again follows \evalplus default setup where the timeout per problem is defined as $T = max(T_{max}, f \times t_{gt})$ with default values of $T_{max} = 1000ms$, $f = 4$ and $t_{gt}$ defined as the measured groundtruth solution time to produce the correct output. 
All timeout related factors can be adjusted to account for variance on different underlying machine and hardware.

\section{Transformation Prompts}

\begin{figure*}[h!]
\centering
\begin{tcolorbox}
[enhanced,size=small,colback=black!5!white,flip title={interior hidden}]
\begin{lstlisting}[captionpos=b, breaklines=true, escapechar=\%]
Here is an example coding problem:
{problem}

Please increase the difficulty of the given coding problem

You can increase the difficulty using the following method:
- Add new constraints and requirements to the original problem, adding approximately 10 additional words.
- Replace a commonly used requirement in the programming task with a less common and more specific one.
- Add more reasoning steps.

Return the new problem in the same format as the example problem (i.e., %
\end{lstlisting}
\end{tcolorbox}
\caption{Prompt for \eedifficult}
\label{fig:difficultprompt}
\end{figure*}

\begin{figure*}[h!]
\centering
\begin{tcolorbox}
[enhanced,size=small,colback=black!5!white,flip title={interior hidden}]
\begin{lstlisting}[captionpos=b, breaklines=true, escapechar=\%]
Here is an example coding problem: 
{problem}

Please generate a more creative coding problem.
You should avoid common programming concepts and instead focus on creating a problem that is interesting and fun to solve.
Return the new problem in the same format as the example problem (i.e., %
\end{lstlisting}
\end{tcolorbox}
\caption{Prompt for \eecreative}
\label{fig:creativeprompt}
\end{figure*}

\begin{figure*}[h!]
\centering
\begin{tcolorbox}
[enhanced,size=small,colback=black!5!white,flip title={interior hidden}]
\begin{lstlisting}[captionpos=b, breaklines=true, escapechar=\%]
Please add a subtle and simple change to the given problem.

You can change the problem using, but not limited to, the following methods:

Add one new requirement to the original problem, such as "Return the list in ascending order", "Return the list in ascending alphabetical order" and "Return unique elements only".

Invert one requirement of the original problem; for instance, reverse the instruction "from shortest to longest" to "from longest to shortest",  reverse "maximum" to "minimum", or reverse "the first" to "the last".

Replace one requirement with another similar but different one; for example, if the original problem requires the values to be sorted, change it to keeping the original order.

Replace constants; for instance, replace zero with one.

Please only apply a minor change, ensuring that the new problem remains logical. Return the new problem in the same format as the original problem (%

Below is the question: 
{problem}
\end{lstlisting}
\end{tcolorbox}
\caption{Prompt for \eesubtle}
\label{fig:subtleprompt}
\end{figure*}

\begin{figure*}[h!]
\centering
\begin{tcolorbox}
[enhanced,size=small,colback=black!5!white,flip title={interior hidden}]
\begin{lstlisting}[captionpos=b, breaklines=true, escapechar=\%]
Here are two example problems:

Example problem 1:
{problem_1}
Example problem 2:
{problem_2}

Please create a new problem that combines problem 1 with problem 2 in a logical way. The new problem should seamlessly integrate the concepts from the two previous examples into a novel context, and require a solution that exercises the understanding of the concepts from both problems. It does not need to take all the inputs from both problems.

An incorrect way would be for the new problem to simply return the answers of the two problems separately. Another incorrect method would be to pass all inputs from both problems but some of them are not used to compute the output.

Return the new problem in the same format as the example problems in %
\end{lstlisting}
\end{tcolorbox}
\caption{Prompt for \eecombine}
\label{fig:combineprompt}
\end{figure*}

\begin{figure*}[h!]
\centering
\begin{tcolorbox}
[enhanced,size=small,colback=black!5!white,flip title={interior hidden}]
\begin{lstlisting}[captionpos=b, breaklines=true, escapechar=\%]
Here is an example coding problem: 
{problem}

Please come up with a new problem which uses helper functions to solve the problem. 

The new problem should contain the following:
first: one or more helper functions 
second: the main problem description consist of the function header and docstring

The main problem description should not refer to the helper function(s) in any way.

The helper function(s) should implement simple parsing or checking logic.

To solve the main problem, one should also use additional complex logic than just calling the helper function(s).

Avoid problems on simple math concepts such as prime, palindrome, anagrams, factorial

Avoid concepts like emails, string or parsing-based problems

Please return the full implementation of the helper function(s) and the main problem description (not the implementation) in the same format as the example problem (%
\end{lstlisting}
\end{tcolorbox}
\caption{Prompt for \eetoolusing}
\label{fig:toolusingprompt}
\end{figure*}

\begin{figure*}[h!]
\centering
\begin{tcolorbox}
[enhanced,size=small,colback=black!5!white,flip title={interior hidden}]
\begin{lstlisting}[captionpos=b, breaklines=true, escapechar=\%]
Below is a coding problem 

{problem}

Make the docstring more verbose and detailed but preserve the semantic meaning
Ensure the function name, input argument names, and example input/output are the same
Return the transformed problem in the same format as the original problem (i.e., function header + docstring)
\end{lstlisting}
\end{tcolorbox}
\caption{Prompt for \eeverbose}
\label{fig:verboseprompt}
\end{figure*}

\begin{figure*}[h!]
\centering
\begin{tcolorbox}
[enhanced,size=small,colback=black!5!white,flip title={interior hidden}]
\begin{lstlisting}[captionpos=b, breaklines=true, escapechar=\%]
Below is a coding problem 

{problem}

Make the docstring shorter and more concise but preserve the semantic meaning
Ensure the function name, input argument names, and example input/output are the same
Return the transformed problem in the same format as the original problem (i.e., function header + docstring)
\end{lstlisting}
\end{tcolorbox}
\caption{Prompt for \eeverbose}
\label{fig:conciseprompt}
\end{figure*}

\begin{figure*}[h!]
\centering
\begin{tcolorbox}
[enhanced,size=small,colback=black!5!white,flip title={interior hidden}]
\begin{lstlisting}[captionpos=b, breaklines=true, escapechar=\%]
Below is a complex coding problem
{problem}

Please decompose the above into 2 smaller sub problems
Return the two modified problems in the same format as the initial problem (i.e., %
\end{lstlisting}
\end{tcolorbox}
\caption{Prompt for \eedecompose}
\label{fig:decomposeprompt}
\end{figure*}

Here we provide the exact targeted transformation prompts used to evolve existing benchmark problems into each of our transformation prompts. 
Figure~\ref{fig:difficultprompt}, ~\ref{fig:creativeprompt}, ~\ref{fig:subtleprompt}, ~\ref{fig:combineprompt}, ~\ref{fig:toolusingprompt}, ~\ref{fig:verboseprompt}, ~\ref{fig:conciseprompt} and ~\ref{fig:decomposeprompt} shows the prompt for \eedifficult, \eecreative, \eesubtle, \eecombine, \eetoolusing, \eeverbose, \eeconcise and \eedecompose respectively. 

\begin{figure*}[h!]
\centering
\begin{tcolorbox}
[enhanced,size=small,colback=black!5!white,flip title={interior hidden}]
\begin{lstlisting}[captionpos=b, breaklines=true, escapechar=\%]
Below is a coding problem
{problem}

Ensure logical coherence in the given problem.
Improve the docstring's clarity and conciseness.
Fix missing or helpful imports.
Include example input/output if absent.
Return the modified problem in the same format as the example problem (i.e., %
\end{lstlisting}
\end{tcolorbox}
\caption{Refinement prompt}
\label{fig:refinementprompt}
\end{figure*}

\begin{figure*}[h!]
\centering
\begin{tcolorbox}
[enhanced,size=small,colback=black!5!white,flip title={interior hidden}]
\begin{lstlisting}[captionpos=b, breaklines=true, escapechar=\%]
Here is a function header with docstring: 
{problem}
Please extract the example raw input argument and expected output from the docstring.
If there are no example input and output please provide new ones.
Return each pair of input and output as assertions in this format:
assert {function_name}({{the_first_input_example}} == {{the_first_output_example}}
assert {function_name}({{the_second_input_example}} == {{the_second_output_example}}
...



Here is a problem with docstring: 
{problem}
Some example inputs and outputs in the docstring may be wrong. Please correct them according to the provided correct assertions below, and ensure that the correct example inputs and outputs %
{assertions}
Return the revised problem in the same format as the original problem (i.e., %
\end{lstlisting}
\end{tcolorbox}
\caption{Input extraction and fixing prompts}
\label{fig:extractionprompt}
\end{figure*}

Figure~\ref{fig:refinementprompt} and ~\ref{fig:extractionprompt} show the refinement and I/O extraction/fixing prompt used in \tech.
The refinement prompt is used to refine the origin generated problem when inconsistency is detected (see Section~\ref{sec:approach}).
The extraction prompt is used to initially obtain a set of testcases from the problem docstring used for self-consistency evaluation. 
We further use an I/O fixing prompt (also in Figure~\ref{fig:extractionprompt}) to fix any examples in the docstring which do not contain the right output (as computed by the groundtruth generated by \gptfour). 

\section{Example Problems in \tech}
\label{sec:exampleprob}

Here we demonstrate a few example problems across the benchmarks in \tech and corresponding \gptfour solution which cannot solve the problem.  
Figure~\ref{fig:difficult_fail}, ~\ref{fig:creative_fail}, ~\ref{fig:subtle_fail}, ~\ref{fig:combine_fail}, ~\ref{fig:tool_using_fail}, ~\ref{fig:verbose_fail} and ~\ref{fig:concise_fail} show such example for the \tech \eedifficult, \eecreative, \eesubtle, \eecombine, \eetoolusing, \eeverbose and \eeconcise respectively. 

\begin{figure*}[h]
\centering
\begin{tcolorbox}
[enhanced,size=small,colback=black!5!white,flip title={interior hidden}]
\begin{lstlisting}[captionpos=b, breaklines=true, language=Python, escapechar=\%]
def common(l1: list, l2: list, n: int):
    """
    Return the n longest strings, sorted by increasing length that are common in two lists.
    However, in the case of a tie, prioritize the string that appears first in list1.
    >>> common(["apple", "banana", "cherry", "date", "elderberry"], ["banana", "date", "fig", "grape", "elderberry"], 3)
    ["banana", "date", "elderberry"]
    >>> common(["cat", "dog", "elephant", "fox", "goat"], ["dog", "fox", "cat", "horse", "iguana", "goat"], 3)
    ["cat", "dog", "goat"]
    >>> common(["hello", "goodbye", "yes", "no", "maybe", "absolutely", "never"], ["no", "maybe", "absolutely", "always", "sometimes"], 1)
    ["absolutely"]
    """
    common_strings = [string for string in l1 if string in l2]
    common_strings.sort(key=len) # does not adjust for tie sorting scenario.
    return common_strings[-n:]
\end{lstlisting}
\end{tcolorbox}
\caption{\gptfour failing solution on problem \tech/58 in \eedifficult}
\label{fig:difficult_fail}
\end{figure*}

\begin{figure*}[h]
\centering
\begin{tcolorbox}
[enhanced,size=small,colback=black!5!white,flip title={interior hidden}]
\begin{lstlisting}[captionpos=b, breaklines=true, language=Python, escapechar=\%]
def crossword_validator(grid: list, words: list):
    """
    Consider a 2D grid representing a square crossword puzzle. Each cell of the grid contains a single character.

    Determine if all the words from a provided list of words can be found in the grid.
    Words can be found left-to-right (in a row), right-to-left (in a row),
    top-to-bottom down (in a column), and bottom-to-top (in a column).

    We assume that words are only found in contiguous cells and words do not 'wrap' around the grid.

    Note: the grid is a square (N x N) and all words are lowercase.

    Parameters:
    grid: list of list of char - where each char is a lowercase letter.
    words: list of str - where each str is a lowercase word.

    Returns:
    boolean - return True all words are found in the grid, False otherwise.

    >>> crossword_validator([['c', 'a', 't'], ['d', 'o', 'g'], ['d', 'o', 'p']], ['cat', 'dog', 'pop'])
    False
    >>> crossword_validator([['c', 'a', 't'], ['d', 'o', 'g'], ['d', 'o', 'p']], ['cdd', 'dog', 'pod'])
    True
    """
    
    n = len(grid)
    for word in words:
        found = False
        for i in range(n):
            # does not correctly check if a word can be found as a sublist of column/grid
            if ''.join(grid[i]) == word or ''.join(grid[i][::-1]) == word: 
                found = True
                break
            if ''.join([grid[j][i] for j in range(n)]) == word or ''.join([grid[j][i] for j in range(n)][::-1]) == word:
                found = True
                break
        if not found:
            return False
    return True
\end{lstlisting}
\end{tcolorbox}
\caption{\gptfour failing solution on problem \tech/21 in \eecreative}
\label{fig:creative_fail}
\end{figure*}

\begin{figure*}[h]
\centering
\begin{tcolorbox}
[enhanced,size=small,colback=black!5!white,flip title={interior hidden}]
\begin{lstlisting}[captionpos=b, breaklines=true, language=Python, escapechar=\%]
from typing import List

def below_one(operations: List[int]) -> bool:
    """
    You're given a list of deposit and withdrawal operations on a bank account that starts with
    one balance. Your task is to detect if at any point the balance of account falls below one, and
    at that point function should return True. Otherwise it should return False.
    >>> below_one([1, 2, 3])
    False
    >>> below_one([1, 2, -4, 5])
    True
    """
    # incorrect starts balance at 0 instead of 1
    balance = 0 
    for operation in operations:
        balance += operation
        if balance < 1:
            return True
    return False
\end{lstlisting}
\end{tcolorbox}
\caption{\gptfour failing solution on problem \tech/3 in \eesubtle}
\label{fig:subtle_fail}
\end{figure*}

\begin{figure*}[h]
\centering
\begin{tcolorbox}
[enhanced,size=small,colback=black!5!white,flip title={interior hidden}]
\begin{lstlisting}[captionpos=b, breaklines=true, language=Python, escapechar=\%]
def grade_with_same_chars(grades: list, s0: str, s1: str):
    """
    A teacher has a special grading system where she assigns her students letter grades
    based on their GPA according to the following table:

             GPA       |    Letter grade
              4.0                A+
            > 3.7                A
            > 3.3                A-
            > 3.0                B+
            > 2.7                B
            > 2.3                B-
            > 2.0                C+
            > 1.7                C
            > 1.3                C-
            > 1.0                D+
            > 0.7                D
            > 0.0                D-
              0.0                E

    However, this year she adds an extra condition: the students can only receive a grade
    if their name (s0) and the course title (s1) have the same characters. The function
    receives a list of GPAs and two strings (s0 and s1) and returns a list of grades
    for the students that pass the condition, or an empty list if no student passes.

    Example:
    grade_with_same_chars([4.0, 3, 1.7, 2, 3.5], 'abcd', 'dcba') ==> ['A+', 'B', 'C-', 'C', 'A-']
    grade_with_same_chars([4.0, 3, 1.7, 2, 3.5], 'aaabcd', 'dddcba') ==> ['A+', 'B', 'C-', 'C', 'A-']
    grade_with_same_chars([4.0, 3, 1.7, 2, 3.5], 'abcd', 'dcbae') ==> []
    """
    # does not correctly check characters are same
    if sorted(s0) != sorted(s1): 
        return []
    grading_system = {4.0: 'A+', 3.7: 'A', 3.3: 'A-', 3.0: 'B+', 2.7: 'B', 2.3: 'B-', 2.0: 'C+', 1.7: 'C', 1.3: 'C-', 1.0: 'D+', 0.7: 'D', 0.0: 'D-', 0.0: 'E'}
    grades_assigned = []
    for gpa in grades:
        for key in sorted(grading_system.keys(), reverse=True):
            if gpa >= key:
                grades_assigned.append(grading_system[key])
                break
    return grades_assigned
\end{lstlisting}
\end{tcolorbox}
\caption{\gptfour failing solution on problem \tech/28 in \eecombine}
\label{fig:combine_fail}
\end{figure*}

\begin{figure*}[h]
\centering
\begin{tcolorbox}
[enhanced,size=small,colback=black!5!white,flip title={interior hidden}]
\begin{lstlisting}[captionpos=b, breaklines=true, language=Python, escapechar=\%]
def is_valid_passport(passport: str) -> bool:
    """ 
    Checks whether a given string is a valid passport number.
    A valid passport number is defined as having exactly 9 alphanumeric characters.

    >>> is_valid_passport('P12345678')
    True
    >>> is_valid_passport('P123456789')
    False
    >>> is_valid_passport('P12345G78')
    False
    """

    if len(passport) != 9:
        return False
    if not passport.isalnum():
        return False
    return True
    
from typing import List, Tuple

def process_passport_data(data: str) -> List[Tuple[str, str]]:
    """
    Parse a string containing passport data and return a list of tuples.
    Each tuple contains the passport number and the name of the passport holder.
    The passport data string is in the format of 'PassportNumber:Name'
    separated by lines.

    Only process valid passport data. A valid passport number is defined as having exactly 9 alphanumeric characters.

    >>> process_passport_data('P12345678:John Doe\nX12345678:Jane Doe')
    [('P12345678', 'John Doe')]
    >>> process_passport_data('P12345678:John Doe\nA123456789:Jane Doe')
    [('P12345678', 'John Doe')]
    """
    valid_passports = []
    lines = data.split('\n')
    for line in lines:
        (passport, name) = line.split(':') # directly splits without checking for validity
        if is_valid_passport(passport):
            valid_passports.append((passport, name))
    return valid_passports
\end{lstlisting}
\end{tcolorbox}
\caption{\gptfour failing solution on problem \tech/23 in \eetoolusing}
\label{fig:tool_using_fail}
\end{figure*}

\begin{figure*}[h]
\centering
\begin{tcolorbox}
[enhanced,size=small,colback=black!5!white,flip title={interior hidden}]
\begin{lstlisting}[captionpos=b, breaklines=true, language=Python, escapechar=\%]
def starts_one_ends(n):
    """
    This function, given a user-provided positive integer 'n', calculates and returns the total count of n-digit positive integers where 1 appears either at the start or the end of the number.

    For instance, if the input is 2, the function will count all the two-digit numbers that either start or end with 1, like 10, 11, 21, 31, ..., 91, and 12, 13, ..., 19.

    It is important to note that this function only accepts positive integers. If a non-integer or non-positive integer is provided, the function may not behave as expected.

    Args:
    n : int
        A positive integer that represents the number of digits in the numbers to be counted.

    Returns:
    int
        The count of the numbers of n-digit positive integers that start or end with 1.

    Examples:
    The function can be called in the following way:
        starts_one_ends(2)
    This example call would return the count of two-digit numbers where 1 appears at the start or at the end.
    """
    if n == 1:
        return 1
    else:
        # incorrectly adds 2 to the final answer
        return 18 * 10 ** (n - 2) + 2
\end{lstlisting}
\end{tcolorbox}
\caption{\gptfour failing solution on problem \humaneval/83 in \eeverbose}
\label{fig:verbose_fail}
\end{figure*}

\begin{figure*}[h]
\centering
\begin{tcolorbox}
[enhanced,size=small,colback=black!5!white,flip title={interior hidden}]
\begin{lstlisting}[captionpos=b, breaklines=true, language=Python, escapechar=\%]
def circular_shift(x, shift):
    """Performs a circular right shift on the digits of x as per shift value and returns the result in string format.
    Returns reversed digits if shift exceeds number of digits.

    >>> circular_shift(12, 1)
    "21"
    >>> circular_shift(12, 2)
    "12"
    """
    x = str(x)
    # does not check shift exceeds string length and return reversed digits
    shift = shift %
    return x[-shift:] + x[:-shift]
\end{lstlisting}
\end{tcolorbox}
\caption{\gptfour failing solution on problem \humaneval/65 in \eeconcise}
\label{fig:concise_fail}
\end{figure*}

\end{document}